\documentclass[12pt, draftclsnofoot, onecolumn]{IEEEtran}
\usepackage{bm,cite,float,amsmath,amssymb,amsthm}
\usepackage{subfigure}
\usepackage[noend]{algpseudocode}
\usepackage{indentfirst}
\usepackage{xcolor}

\usepackage[linesnumbered,boxed,ruled,commentsnumbered]{algorithm2e}
\usepackage{algpseudocode}  
\usepackage{amsmath}  
\floatname{algorithm}{algorithm}
\renewcommand{\algorithmicrequire}{\textbf{Input:}}  % Use Input in the format of Algorithm  
\renewcommand{\algorithmicensure}{\textbf{Output:}} % Use Output in the format of Algorithm

\makeatletter
\def\BState{\State\hskip-\ALG@thistlm}
\makeatother

\usepackage{amssymb}
\usepackage{amsmath}
\usepackage{graphicx}
\usepackage{cite}
\usepackage{balance}
\usepackage[utf8]{inputenc}

\IEEEoverridecommandlockouts

\usepackage{graphicx,epstopdf}
\usepackage{epsfig}	
\usepackage{amsfonts,balance}
\usepackage{bbm}

\usepackage[justification=centering]{caption}

\floatname{algorithm}{Algorithm}
\usepackage{lipsum}

\newtheorem{theorem}{Theorem}

\newtheorem{corollary}{Corollary}

\makeatletter
\def\ScaleIfNeeded{%
\ifdim\Gin@nat@width>\linewidth \linewidth \else \Gin@nat@width
\fi } \makeatother

\begin{document}

\title{A General Deep Reinforcement Learning Framework for Grant-Free NOMA  Optimization in mURLLC
}

%Deep Reinforcement Learning-based Grant-Free Non-Orthogonal Access for  URLLC Service

\author{ Yan Liu, ~\IEEEmembership{Student Member,~IEEE,}
		Yansha Deng,
		~\IEEEmembership{Member,~IEEE,}\\
			Hui Zhou,
		~\IEEEmembership{Student Member,~IEEE,}
		 Maged Elkashlan,~\IEEEmembership{Senior Member,~IEEE,}\\
		 and Arumugam Nallanathan, ~\IEEEmembership{Fellow,~IEEE}

\thanks{Y. Liu is with the Key Laboratory of Ministry of Education in Broadband
Wireless Communication and Sensor Network Technology, Nanjing University
of Posts and Telecommunications, Nanjing 210003, China and with School of Electronic Engineering and Computer Science, Queen Mary University of London, London, UK
(e-mail:\{yan.liu\}@qmul.ac.uk). }
\thanks{M. Elkashlan, and A. Nallanathan are with School of Electronic Engineering and Computer Science, Queen Mary University of London, London, UK
(e-mail:\{yan.liu,  maged.elkashlan, a.nallanathan\}@qmul.ac.uk). }
\thanks{ Y. Deng and H. Zhou are with Department of Engineering, King's College London, London, UK (e-mail:\{yansha.deng, hui.zhou\}@kcl.ac.uk).)
(Corresponding author: Yansha Deng (e-mail:\{yansha.deng\}@kcl.ac.uk).)}
\vspace*{+0.3cm}
}

\maketitle

\vspace*{-2cm}

\begin{abstract}
Grant-free non-orthogonal multiple access (GF-NOMA) is a potential technique to support massive Ultra-Reliable and Low-Latency Communication (mURLLC) service.  However,  the dynamic resource configuration in GF-NOMA systems is challenging due to random traffics and collisions, that are unknown at the base station (BS). Meanwhile, joint consideration of the latency and reliability requirements makes the resource configuration of GF-NOMA for mURLLC more complex. To address this problem, we develop a general learning framework for signature-based GF-NOMA in mURLLC service taking into account the multiple access signature collision, the UE detection, as well as the data decoding procedures for the K-repetition GF and the Proactive GF schemes.  The goal of our learning framework is to maximize the long-term average number of successfully served users (UEs) under the latency constraint. We first perform a real-time repetition value configuration based on a double deep  Q-Network  (DDQN) and then propose a  Cooperative  Multi-Agent  learning technique based on the DQN (CMA-DQN) to optimize the configuration of both the repetition values and the contention-transmission unit (CTU) numbers. 
Our results show that the number of successfully served UEs  under the same latency constraint in our proposed learning framework is up to ten times for the K-repetition scheme, and two times for the Proactive scheme, more than that with fixed repetition values and CTU numbers.
In addition, the superior performance of CMA-DQN over the conventional load estimation-based approach  (LE-URC) demonstrates its capability in dynamically configuring in long term.
Importantly, our general learning framework can be used to optimize the resource configuration problems in all the signature-based GF-NOMA schemes.

\end{abstract}

\begin{IEEEkeywords}
mURLLC, NOMA, grant free,  deep reinforcement learning, resource configuration.
\end{IEEEkeywords}

\maketitle

\section{Introduction}

%In Rel-15, the UL transmission in RRC connected mode with configured-grant transmission is specified as grant-free scheme. 

As a new and dominating service class in 6th Generation (6G) networks, massive Ultra-Reliable and Low Latency Communications (mURLLC) integrates URLLC with massive access  to support massive short-packet data communications in time-sensitive wireless networks with  high  reliability and low access latency\cite{9186090}. This requires a reliability-latency-scalability trade-off and mandates a principled and scalable framework accounting for the delay, reliability,   and decision-making under uncertainty \cite{8869705}.
Concretely speaking, the Third Generation Partnership Project  (3GPP) standard \cite{3gpp2018study} has defined a general URLLC requirement  as: $1-10^{-5}$ reliability within 1ms user plane latency
for 32 bytes.
More details on the requirements of various different URLLC use cases, including smart grids, intelligent transportation systems, and process automation with reliability requirements of $1-10^{-3}$ to $1-10^{-6}$ at latency requirements between 1 ms to 100 ms, can be found in \cite{3gpp22261}. It is also anticipated that the device density may grow to hundred(s) of devices per cubic meter in the 6G white paper\cite{latva2020key}.

\begin{figure}[htbp!]
	\centering
	\includegraphics[width=5.6in,height=2.5in]{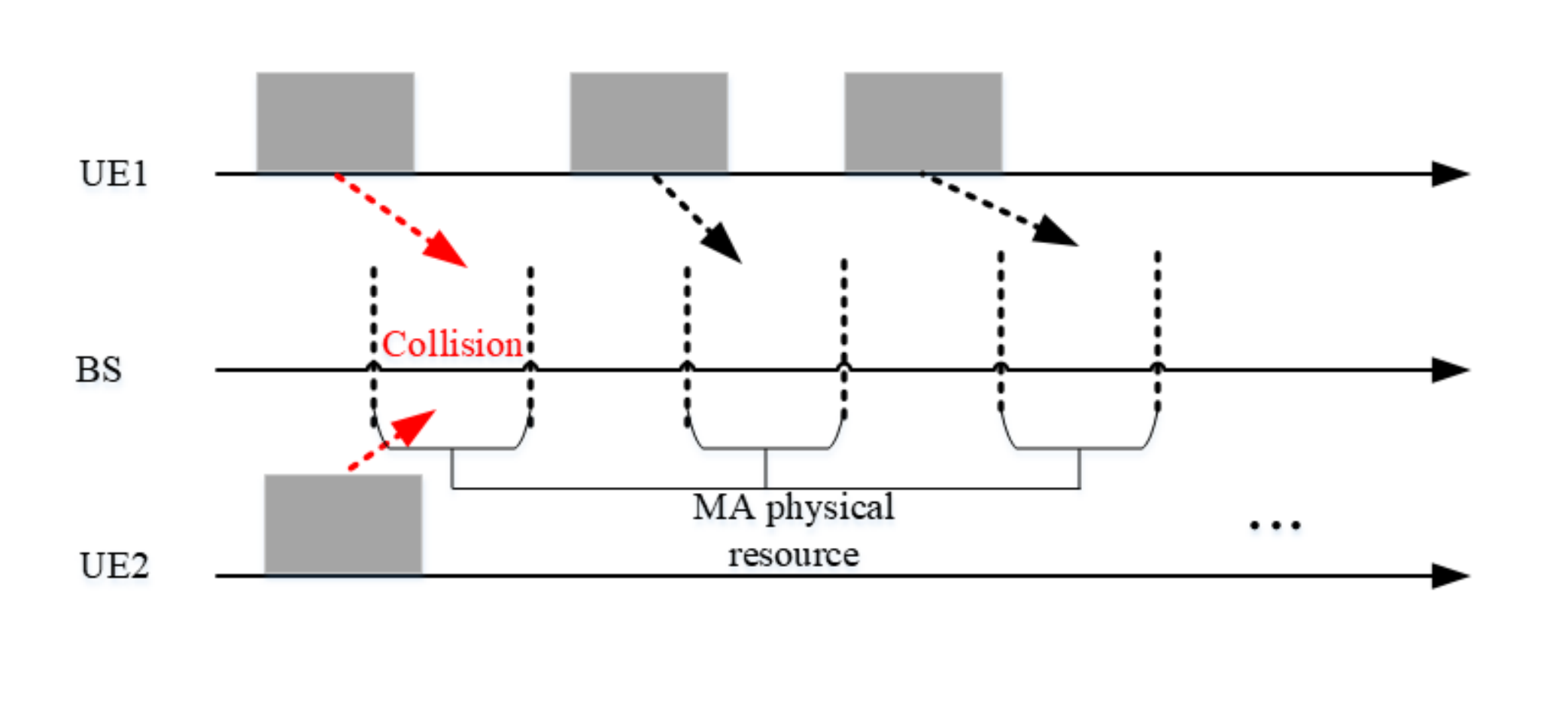}
	\caption{Collision of MA physical resource.
	}
	\label{fig:1}
\end{figure}

Current cellular network can hardly  fulfill  the joint massive connectivity, ultra-reliability, and low latency requirements in mURLLC service.
To achieve low latency, \textit{grant-free (GF) access} has been proposed\cite{11705654,ye2018uplink} as an alternative for traditional grant-based (GB) access due to its drawbacks in high latency and heavy signaling overhead\cite{6163599}.
Different from GB access,  GF access allows a User Equipment (UE) to transmit its data  to the Base Station (BS) in an arrive-and-go manner, without sending a scheduling request (SR) and obtaining a resource grant (RG) from the network\cite{R1-1608917}.
To achieve high reliability, several GF schemes, including the \textit{K-repetition} scheme and the \textit{Proactive} scheme, have been proposed, where a pre-defined number ($K$) of consecutive replicas of the same packet are transmitted \cite{3gpp38214,1705246,1612246,1903079}.
To achieve massive connectivity, \textit{non-orthogonal multiple access (NOMA)} has been proposed to synergize with GF in order to deal with the multiple access (MA) physical resource collision as shown in Fig.~\ref{fig:1} in contention-based\footnote{Unless otherwise stated, the GF and GB access described in this work are both contention based.} GF access on orthogonal multi-access (OMA) physical resources, when two or more UEs transmit their data using the same MA physical resource\cite{8533378,9097306}.
Here, we focus on the  signature-based GF-NOMA, including the sparse code multiple access (SCMA), multiuser shared access (MUSA),  pattern division multiple access (PDMA), and etc, where the NOMA technique allows multiple users to transmit over the same MA physical resource by employing user-specific signature patterns (e.g, codebook, pilot sequence, interleaver/mapping pattern, demodulation reference signal, power, etc.)\cite{7972955}.
However,  when two or more UEs transmit their data using the same MA physical resource and the same MA signature, the   MA  signature collision  occurs, and the BS cannot differentiate among different UEs and therefore cannot decode the data\cite{8533378,9097306}.

It is important to know that the research challenges in  GF-NOMA are fundamentally different from those in GB-NOMA \cite{9210822,8408843}.
\begin{figure}[htbp!]
	\centering
	\includegraphics[width=5.6in,height=2.6in]{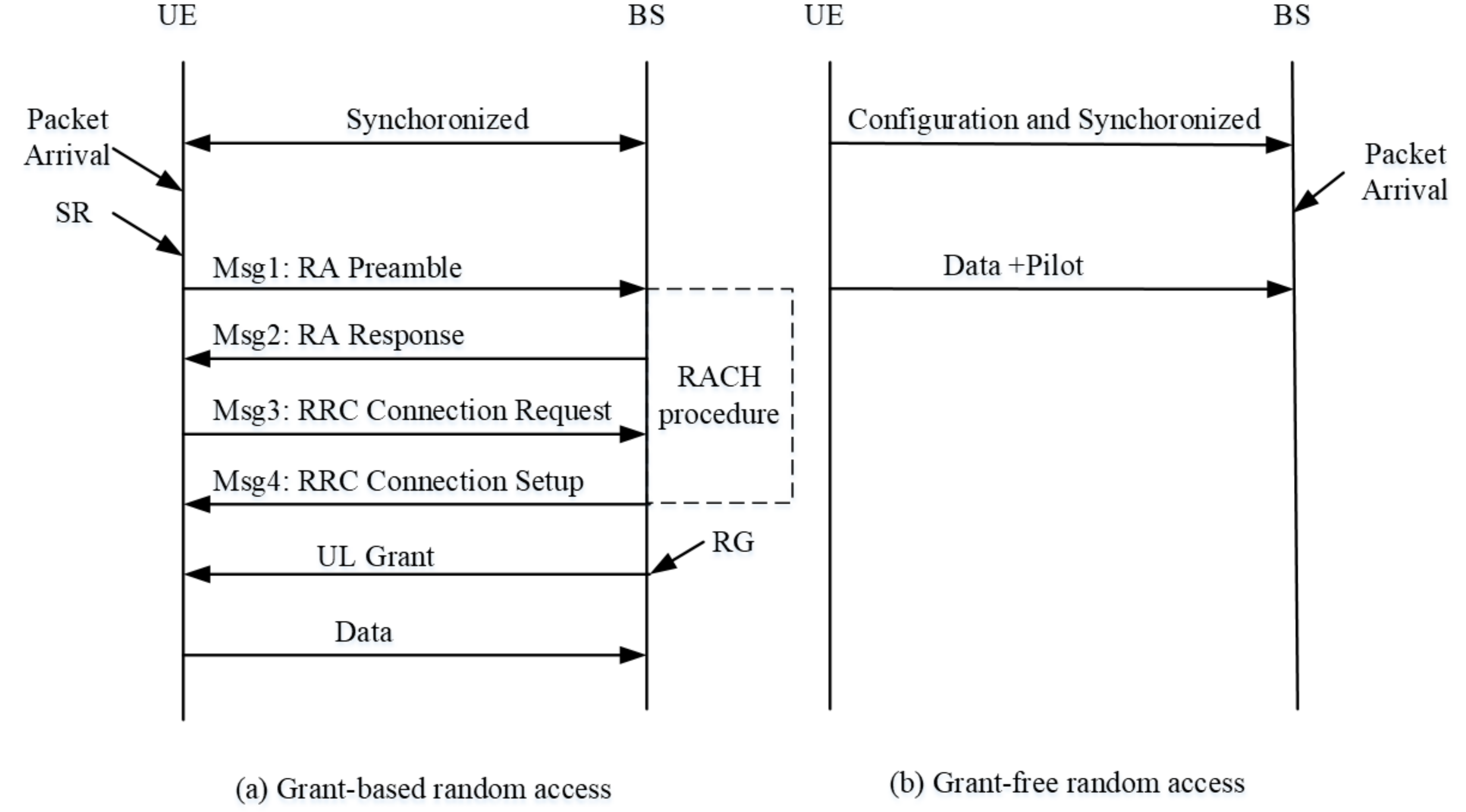}
	\caption{Uplink transmissions for grant-based and grant-free random access.
	}
	\label{fig:2}
\end{figure}
In GB scheme, the four-step random access (RA) procedure as shown in Fig.~\ref{fig:2} is executed by the UE to request the BS to schedule dedicated resources  for data transmission, where the data transmission is more likely to be successful once the random access succeeds. 
While in GF scheme, the data is transmitted along with the pilot in a randomly chosen MA resource, which is unknown at the BS, and can lead to new research problems including but not limited to: 1)  identify  the MA signature collisions due to UEs transmitting data over the same channel resource with the same MA signature; 2) blind detection of active UEs due to that the set of active users is unknown to the BS;  3) blind decoding of UEs’ data with no knowledge of  channels and codebooks.
% In GF-NOMA, blind detection is crucial to the performance of communication. However, how to do blind detection and based on what to detect is the major problem that needs to be investigated with this approach.
% One option is to use pilots, as was done in GF-SCMA \cite{7063547}, where pilots may serve the purpose of both the UE activity detection and channel estimation. 
These new research problems make the dynamic resource configuration of GF-NOMA as one important but challenging problem to solve.

%Though the GF-NOMA technique increases the resource utilization efficiency, the MA physical resources and the pilots are still limited due to the dramatically  increasing  number  of  UEs over years. 
The main challenges of the dynamic resource configuration optimization of GF-NOMA include:
1) the set of active users and their respective channel conditions are unknown to the BS, which prohibits the pre-configuration and the pre-assignment of resources, including  pilots/preambles, power, codebooks, repetition values, retransmission  related  parameters, and etc;
2) simultaneously satisfy  the reliability and latency requirements  under  random traffics, the optimal parameter configurations vary over different time slots, which is hard to be described  by a tractable  mathematical model;
3) the MA signature collision detection and the blind UE activity detection, as well as the data decoding, need to be considered, which largely impacts the resource configuration in each time slot;
4) a general optimization framework for GF-NOMA systems have never been established for various signature-based NOMA schemes.
%These parameters could also change randomly from one time slot to the next. In addition, 
%most works focus on user detection and correct estimation of parameters, which assume that users have unique signature sequences and thus collisions are not an issue. However, 

The above mentioned challenges can hardly be solved via the traditional convex optimization method, due to the complex communication environment with the lack of tractable mathematical formulations, whereas Machine Learning (ML),  can be a potential alternative approach, due to that it solely relies on the self-learning of the environment interaction, without the need to derive explicit optimization solutions based on a complex mathematical model.
In the GF-NOMA system,  the BS can only observe the results of both collision detection (e.g., the number of non-collision UEs and collision MA signatures) and data decoding (e.g., the number of successful decoding UEs and failure decoding UEs) in each round trip time (RTT). 
This historical information can be potentially used to facilitate the long-term optimization of future configurations.
Even if one knew all the relevant statistics, tackling this problem in an exact manner would result in a Partially Observable Markov Decision Process (POMDP) with large state and action spaces, which is generally intractable. 
To deal with it, Reinforcement Learning (RL), can be a promising tool to deal with this complex POMDP problem in GF-NOMA resource
configuration optimization.
%The complexity of the problem is compounded by the lack of a prior knowledge at the BS regarding the stochastic traffic and unobservable channel statistics (i.e., random collision, and effects of physical radio including path-loss as well as fading).

In this paper, we aim to develop  a general learning framework for GF-NOMA systems with mURLLC services.
%To showcase the efficiency, we compare the proposed RL-based approaches with the conventional heuristic uplink resource allocation approaches. 
Our contributions can be summarized as follows:

\begin{itemize}
    \item We develop a general learning framework for  dynamic resource configuration optimization in signature-based GF-NOMA systems, including the SCMA, MUSA, PDMA, and etc, for mURLLC service.
In this framework, we practically simulate the random traffics, the resource selection and configuration, the transmission
latency check, the collision detection, the data decoding, and the Hybrid Automatic Repeat reQuest (HARQ) retransmission procedures.
We use this generated simulation environment to train the RL agents.
\end{itemize}

\begin{itemize}
    \item  
    We first perform the repetition values dynamic optimization via developing a double Deep Q-Network (DDQN) to optimize the number of successfully served UEs under the latency constraint for the K-repetiton GF scheme and the Proactive GF scheme, respectively. 
We then develop a Cooperative Multi-Agent learning based on DQN (CMA-DQN) to dynamically optimize both the repetition values and contention-transmission unit  (CTU)  numbers, which breaks down the selection in high-dimensional parameters into multiple parallel sub-tasks with a number of DQN agents cooperatively being trained to produce each parameter.
\end{itemize}

\begin{itemize}
    \item Through dynamic optimization via CMA-DQN, we show that the Proactive scheme outperforms the K-repetition scheme in terms of the number of successfully served UEs, especially under the long latency constraint, which are opposite to the results  without optimization in our previous work\cite{9174916} with only a single packet transmission, 
% Our results  shown that the number of successfully served UEs under the same latency constraint in our proposed CMA-DQN framework is up to ten times for the K-repetition scheme, and two times for the Proactive scheme, more than that with fixed repetition values and CTU   numbers, respectively.
 Our  results  also show the superior performance of CMA-DQN over the conventional load estimation-based approach  (LE-URC) in the multiple configuration parameters scenario.
%  that  in the multiple configuration parameters scenario, the superiority  of  the  proposed  learning  framework  over  the  conventional  load  estimation-based  approach (LE-URC)  is  significant.
%  In addition, the superior performance of CMA-DQN over the conventional load estimation-based approach  (LE-URC) demonstrates its capability in dynamically configuring in long term.
 Importantly,  our  general  learning  framework  can  be  used  to  optimize  theresource configuration problems in all the signature-based GF-NOMA schemes.
 %Our proposed  learning framework  can be extended to optimize the resource configuration problems in other GF-NOMA schemes.
\end{itemize}

The rest of the paper is organized as follows. 
Section II presents related works. Section III illustrates the system model and formulates the problem.
Section IV illustrates the
preliminary and the conventional approach.
Section V proposes
Q-learning based uplink GF-NOMA resource configuration approaches.
Section VI elaborates
the numerical results, and finally, Section VII summarizes the
conclusion and future work.

\section{Related Works}

The potential of GF-NOMA in diverse services  is an open research area as the research challenges of GF-NOMA have not been solved yet.
In \cite{7887683,8352623,8288402}, GF-NOMA
is designed empirically by directly incorporating the GF mechanism into several state-of-the-art NOMA schemes, including SCMA, MUSA, and PDMA, that are categorized according to their specially designed spreading signatures.
Most existing  GF-NOMA works focused on  the receiver design using the compressive sensing (CS) technique.
The authors in \cite{7731144} proposed a message passing algorithm  to solve the problem of GF-NOMA using CS-based approaches, which improves the 
bit error rate (BER) performance in comparison to \cite{7462187}.
The authors in \cite{7996932} considered a comprehensive study
such that synchronization, channel estimation, user detection
and data decoding were performed in one-shot. The proposed
CS-based algorithm exploited  the sparsity of
the system in terms of user activity and multi-path fading.
To the best of our knowledge, no works have focused on the optimization of a general GF-NOMA system taking into account  the MA collision detection, the UE detection, as well as data decoding procedures. 
The major challenge comes from the fact that random user activation and non-orthogonal transmissions make the GF-NOMA difficult to mathematically model and optimize using traditional optimization methods.

Machine Learning has been introduced  to improve the GF-NOMA systems in \cite{8625480,8952876,8968401}.
In \cite{8625480}, deep learning was used to solve a variational
optimization problem for GF-NOMA. 
The neural network model includes encoding, user activity, signature sequence generation, and decoding. 
The authors then extended their work 
to design a generalized/unified framework for NOMA using
deep multi-task learning in \cite{8952876}. 
A deep learning-based active user detection scheme has been proposed for GF-NOMA in \cite{8968401} . 
By feeding training data into the designed deep neural network, the proposed active user detection scheme learns the nonlinear mapping between received NOMA signal and indices of active devices.
Their results shown that the trained network can handle the whole
active user detection process, and achieve accurate detection
of the active users. 
These works assumed that each UE is pre-allocated with a unique sequence, and thus collisions are not an issue. However, this assumption does not hold in massive UEs settings in mURLLC, where the collision is the bottleneck of the GF-NOMA performance. 
Different from \cite{8625480,8952876,8968401}, we aim to develop a general learning framework to optimize  GF-NOMA systems for  mURLLC service taking into account the  MA signature collision, the UE detection as well as the data decoding procedures.

\section{System Model}

We consider a single cell network consisting of  a BS  located at the center and a set of $N$ UEs randomly located in an area of the plane $\mathbb{R}^2$, where the UEs are in-synchronized and unaware of the status of each other.
Once deployed, the UEs remain spatially static.
The time is divided into short transmission time intervals (TTIs)\footnote{In this paper, the TTI refers to a mini-slot. The Fifth Generation (5G) New Radio (NR) introduces the concept of `mini-slots' and supports a scalable numerology allowing the sub-carrier spacing (SCS) to be expanded up to 240 kHz. In contrast with the LTE slot consisting of 14 OFDM symbols per TTI, the number of OFDM symbols in 5G NR mini-slots ranges from 1 to 13 symbols, and the larger SCS decreases the length of each symbol further. Collectively, mini-slots and flexible numerology  allow shorter transmission slots to meet the stringent latency requirement. }, 
and the small packets for each UE are generated according to random inter-arrival processes over the short-TTIs, which are Markovian as defined in \cite{name2015,3gpp37868} and unknown to BS.
%For each UE, there is a random process for the generation of uplink small data packets, which is unknown to the BS.
%We consider the time is divided into short-TTIs\footnote{5G NR introduces the concept of `mini-slots' and supports a scalable numerology allowing the sub-carrier spacing (SCS) to be expanded up to 240 kHz. In contrast with the LTE slot duration of 14 OFDM symbols per TTI, mini-slots in 5G NR can compose of 1-13 symbols.  Collectively, this allows shorter transmission slots to meet the stringent latency requirement.
%In this paper, the TTI refers to a mini-slot.}, and the packets inter-arrival processes are independent and
%identically distributed over the TTIs, which are Markovian as defined in \cite{name2015}\cite{3gpp37868}.
\begin{figure}[htbp!]
	\centering
	\includegraphics[width=6.6 in,height=3in]{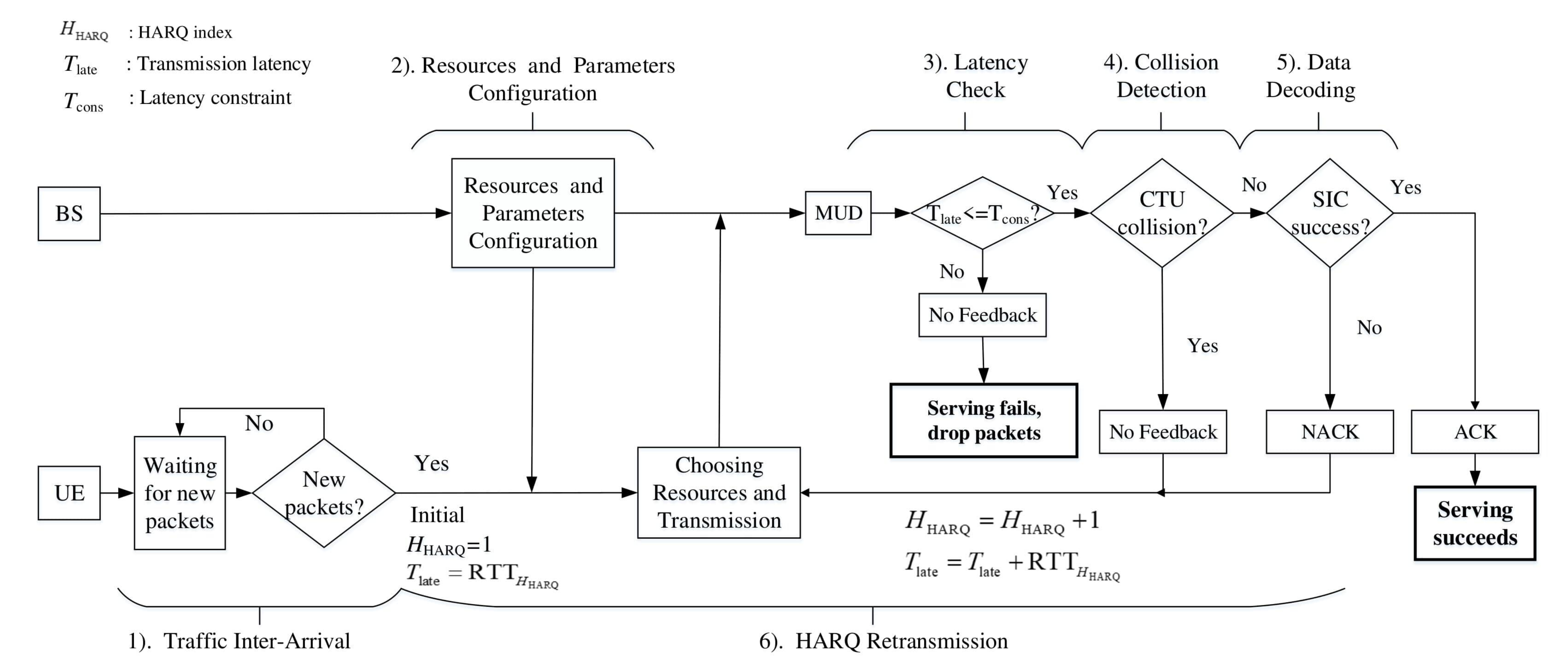}
	\caption{Uplink GF-NOMA transmission procedure.
	}
	\label{fig:3}
\end{figure}
\subsection{ GF-NOMA Network Model}
We consider the uplink contention-based GF-NOMA  over a set of preconfigured MA resources.
To capture the effects of the physical radio,
we consider the standard power-law path-loss model with the path-loss attenuation $r^{-\eta}$, where $r$ is the Euclidean distance between the UE and the BS and $\eta$ is the path-loss attenuation factor. 
In addition, we consider a Rayleigh flat-fading environment, where the channel power
gains $h$ are exponentially distributed (i.i.d.) random variables with unit mean.
We presents the uplink GF-NOMA procedure in Fig.~\ref{fig:3} following the 3GPP standard \cite{Solutions,R1-1608917,MA,1612246,1903079}, which includes 1) traffic inter-arrival, 2) resources and parameters configuration, 3) latency check;
4) collision detection, 5) data decoding, and 6) HARQ retransmissions.
These six stages are explained in details in  the following six subsections to illustrate  the system model.
%According to \cite{nan2018collision}\cite{8533378}, each UE transmits its pilot and data simultaneously.
%Each UE has a data buffer that stores packets received from higher layers.
%As only active UEs will try to request for uplink channel resources, we define the active probability of each IoT UE pa ∈[0, 1] follows a Bernoulli process.
%An i.i.d. Bernoulli traffic generation model with probability of $p_a\in [0, 1]$, is assumed at each buffer.
%At each time slot, transmitters with non-empty buffers employ the F-ALOHA protocol with probability pfa  pa/Nc to access one of the Nc channels, where pa is the ALOHA transmission probability.
\subsubsection{Traffic Inter-Arrival}
We consider a bursty traffic scenario, where massive UEs are triggered due to an emergency event, e.g., earthquake alarm and fire alarms\cite{3gpp37868,8985528}. Each UE would be activated at any time $\tau$, according to a time limited Beta probability density function as \cite[Section 6.1.1]{3gpp37868}
\begin{align}\label{beta}
p(\tau) = \frac{{{\tau ^{\alpha  - 1}}{{(T - \tau )}^{\beta  - 1}}}}{{{T^{\alpha  + {\beta}  - 1}} \rm {Beta}(\alpha ,\beta )}},
\end{align}
where  $T$ is the total time of the bursty traffic and Beta ($\alpha , \beta$) is the Beta function with the constant
parameters $\alpha$ and $\beta$\cite{gupta2004handbook}. 

Due to the nature of slotted-Aloha, a UE can only transmit at the beginning of a RTT as shown in Fig. 4 and Fig. 5, which means that the newly activated UEs executing transmission come from those who received an packet within the interval between with the last RTT period ($\tau^{i-1}$,$\tau^{i}$).
The traffic instantaneous rate in packets in a period is described by a function $p(\tau)$, so that the packets arrival rate in the $i$th RTT  is given by
\begin{align}\label{rate}
    {\mu ^i} = \int_{{\tau _{i - 1}}}^{{\tau _i}} {p(\tau )} d\tau.
\end{align}

\subsubsection{Resources  and Parameters  Configuration} 
\begin{figure}[htbp!]
	\centering
	\includegraphics[width=5.3 in,height=2.2in]{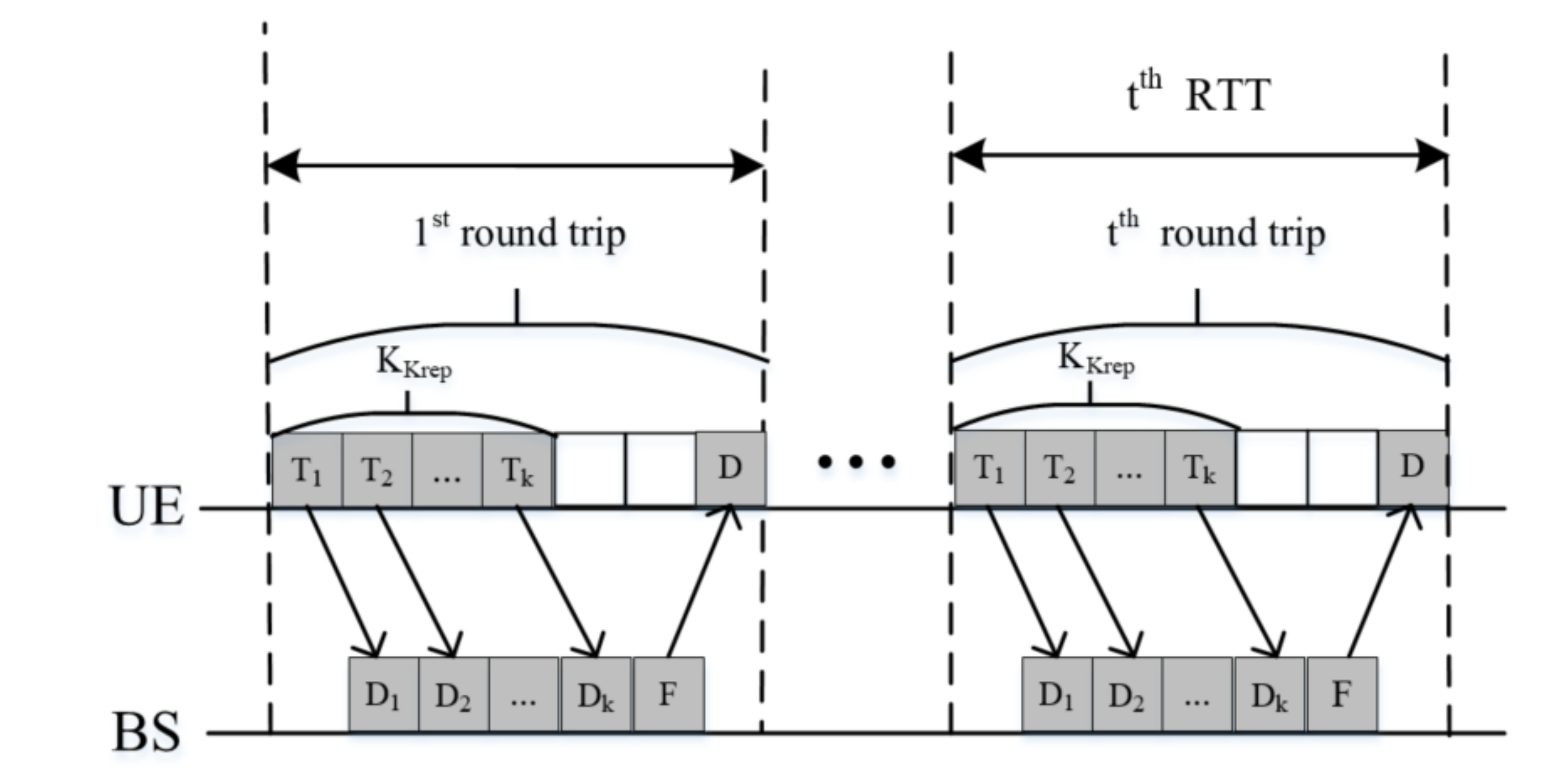}
	\caption{K-repetition GF transmission
	}
	\label{fig:4}
\end{figure}

The  MA resources,
repetition values, and HARQ related parameters, etc, are configured at the BS  by radio resource control (RRC) signaling and L1 signaling
prior to the GF access (as Type 2 GF \cite{3gpp38824}).
%Once the data of a UE arrives, the UE randomly selected one of the configured MA resources to transmit its data under a GF scheme.
%transmitted immediately using a randomly selected MA resource under a repetition scheme.

a) Repetition values: 
We consider the K-repetition GF scheme and the Proactive GF scheme  as shown in Fig.~\ref{fig:4} and Fig.~\ref{fig:5}, respectively.
%Each scheme has its own configured repetition value, i.e., $K_{Krep}$ for K-repetition scheme and $K_{Proa}$ for Proactive scheme.
The repetition values  for K-repetition scheme $K_{\rm Krep}^t$ and for Proactive scheme  $K_{\rm Proa}^t$ are configured at the beginning of each RTT.

\begin{itemize}
\item \textbf{K-{repetition} scheme:}
The K-repetition scheme is illustrated in Fig.~\ref{fig:4}, the UEs served by the BS are configured to autonomously transmit (T) the same packet for $K_{\rm Krep}^t$ repetitions in consecutive TTIs. 
The BS decodes (D) each repetition independently and the transmission in one RTT is successful when at least one repetition succeeds.
After processing all the received $K_{\rm Krep}^T$ repetitions, the BS transmits the ACK/NACK feedback (F) to the UE.
\begin{figure}[htbp!]
	\centering
	\includegraphics[width=5.6 in,height=2.6in]{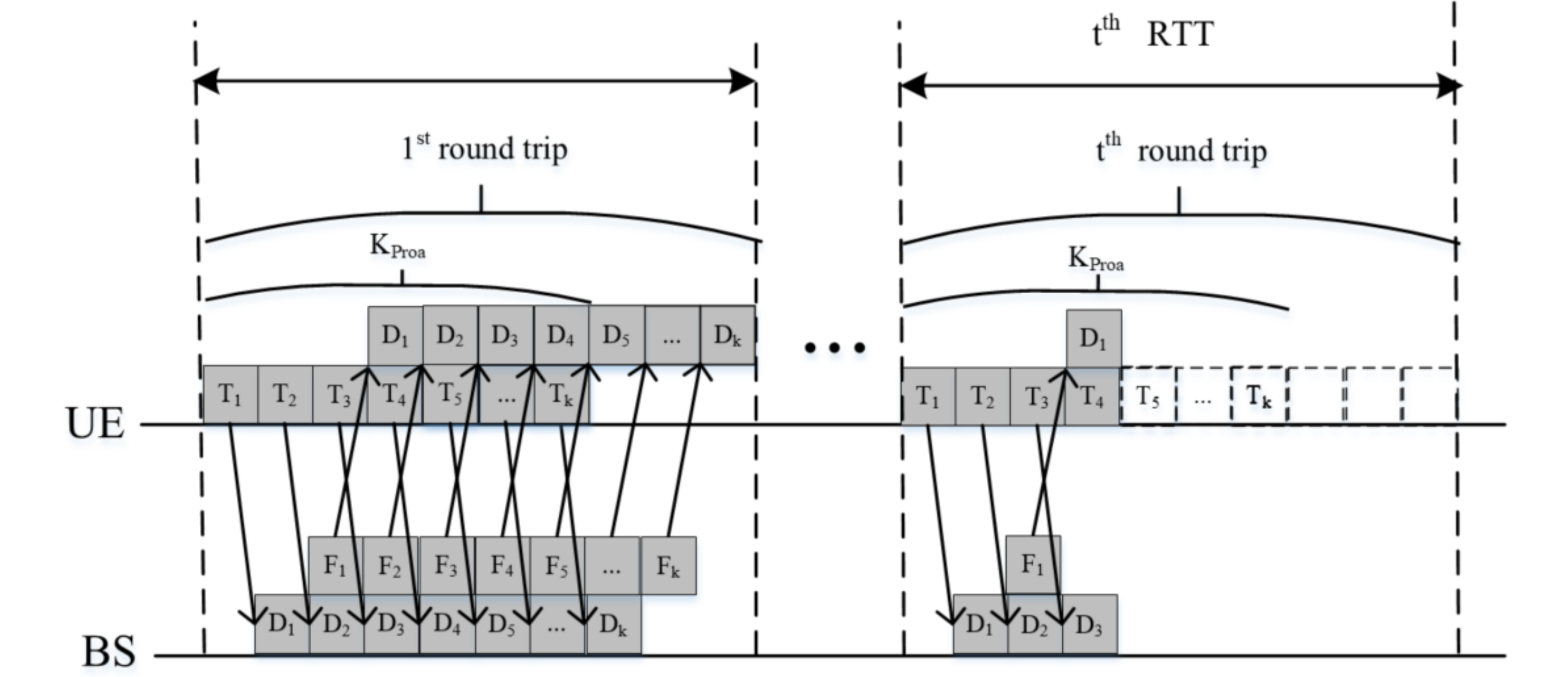}
	\caption{Proactive GF transmission
	}
	\label{fig:5}
\end{figure}
\item \textbf{Proactive scheme:}
The Proactive scheme is illustrated in Fig.~\ref{fig:5}. Similar to the K-repetition scheme, the UEs served by the BS are configured to repeat the transmission for a maximum number of $K_{\rm Proa}^t$ repetitions, but can receive the feedback after each repetition. This allows the UE to terminate repetitions earlier once receiving the ACK. 
\end{itemize} 

\begin{figure}[htbp!]
	\centering
	\includegraphics[width=6.2 in,height=2.4in]{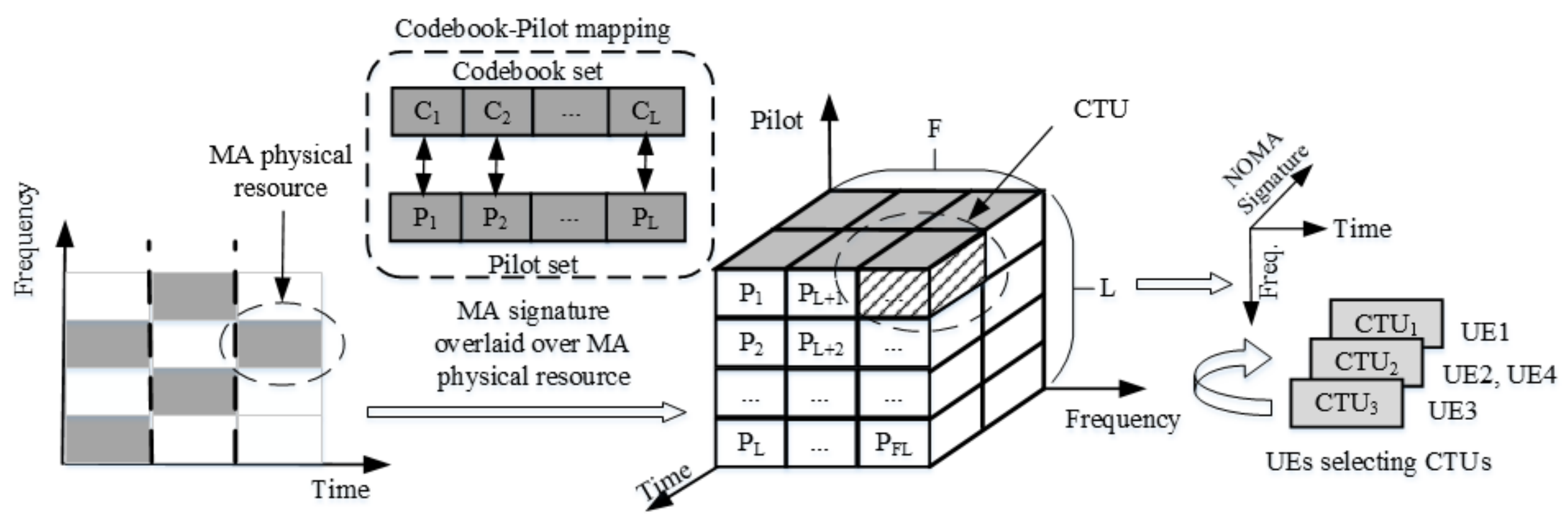}
	\caption{GF-NOMA resource
	}
	\label{fig:6}
\end{figure}

Considering the small packets of mURLLC traffic, we set the packet transmission  time as one TTI. The BS feedback time   and the BS (UE) processing time are also assumed to be one TTI following  our previous work \cite{9174916}.
Once the repetition value is configured, the duration of one RTT is known to the UEs and the BS, which is given as
\begin{align}\label{RTT}
T_{{\rm RTT}}^t=(K^t+3) {\rm TTIs}, 
\end{align}
with $K^t=K_{\rm Krep}^t$ or $K^t=K_{\rm Proa}^t$ for the K-repetition scheme and the Proactive scheme, respectively.

 b) MA resources:
%Once activated at a round trip, each UE  pre-configured resources and  are known in advance by UEs as the basic units for contention. 
A \textit{contention-transmission
unit} (CTU) as shown in Fig.~\ref{fig:6} is defined as the basic MA resource, where each CTU may comprise of
%several fields including radio resources, reference signal,nd spreading sequence
a MA physical resource and a MA signature\cite{ye2018uplink,9097306,7063547}.
The MA physical resources  represent a set of time-frequency resource blocks (RBs). 
The MA signatures represent a set of pilot sequences for channel estimation and/or UE activity detection, and a set of codebooks  for robust data transmission and interference whitening, etc.
Without loss of generality, in one TTI, we consider $F$ orthogonal RBs and each RB is overlaid with $L$ unique codebook-pilot\footnote{A one-to-one mapping or a many-to-one mapping between the pilot sequences  and codebooks can be predefined.
Since it  has been verified in \cite{7887683} that the performance loss due to codebook collision is negligible for a real system, we focus on the pilot sequence collision and consider the one-to-one mapping as \cite{8533378,9031550}. %Thus, the UE selects the unique codebook means that it selects the unique pilot sequence. Thus, the UE selects the unique codebook means that it selects the unique pilot sequence. We will use the 'codebook' throughout the paper.
} pairs\cite{8533378,8316582}.
%To enable NOMA, in each RB, there are $L$ codebooks  and each coodbook associates its unique pilot sequences\footnote{A one-to-one mapping or a many-to-one mapping between the pilot sequences  and codebooks can be predefined.
%Since it is described in \cite{7887683} that the performance loss due to codebook collision is negligible for a real system, we focus on the pilot sequence collision and consider the one-to-one mapping in our work like \cite{8533378}\cite{9031550}. %Thus, the UE selects the unique codebook means that it selects the unique pilot sequence. Thus, the UE selects the unique codebook means that it selects the unique pilot sequence. We will use the 'codebook' throughout the paper.
%} \cite{8533378}\cite{8316582}.
Thus, at the beginning of each RTT, the BS configures a resource pool of  $C^t=F\times L$ unique CTUs, and each UE randomly choose one CTU from the pool to transmit in this RTT.

\subsubsection{Latency Check}
The HARQ index $H_{HARQ}$ is included in the pilot sequence and can be detected by the BS. At the beginning of each RTT, the HARQ index and the transmission latency $T_{\rm late}$ will be updated as shown in Fig.~\ref{fig:3}.
For example, for the initial RTT with initial $K^1$, $H_{HARQ}=1$ and $T_{\rm late}=RTT_{H_{HARQ}=1}$, where $RTT_{H_{HARQ}}$ is calculated by using \eqref{RTT}. 
After this round time trip transmission, the BS optimizes a $K^2$ based on the observation of the reception and configures it to the UE for the next RTT. Then, the UE updates its $H_{HARQ}=2$ and calculated $RTT_{H_{HARQ}=2}$ by using \eqref{RTT} with $K^2$, and consequently, the transmission latency $T_{\rm late}$ is updated as $T_{\rm late}=RTT_{H_{HARQ}=1}+RTT_{H_{HARQ}=2}$.
When $T_{\rm late}>T_{\rm cons}$, the UE fails to be served and the packets will be dropped. Note that the HARQ index, as well as the transmission latency, will be updated at the beginning of each RTT instead of at the end of each RTT due to that we consider the user plane latency\footnote{User plane latency is defined as the one-way latency from the processing of the packet at the transmitter to the successful reception of the packet, including  the transmission processing time, the transmission time, and the reception processing time.} in this work. That is to say, from the UE perspective, when the UE executes this RTT, it will check transmission results after finishing the RTT. Thus, the duration of this RTT should be included when calculating the UE transmission latency.

\subsubsection{Collision Detection}
At each RTT, each active UE transmits its packets to the BS by randomly choosing a CTU.
The BS  can  detect the UEs that have chosen different CTUs.
\begin{figure}[htbp!]
	\centering
	\includegraphics[width=5.3in,height=3.5in]{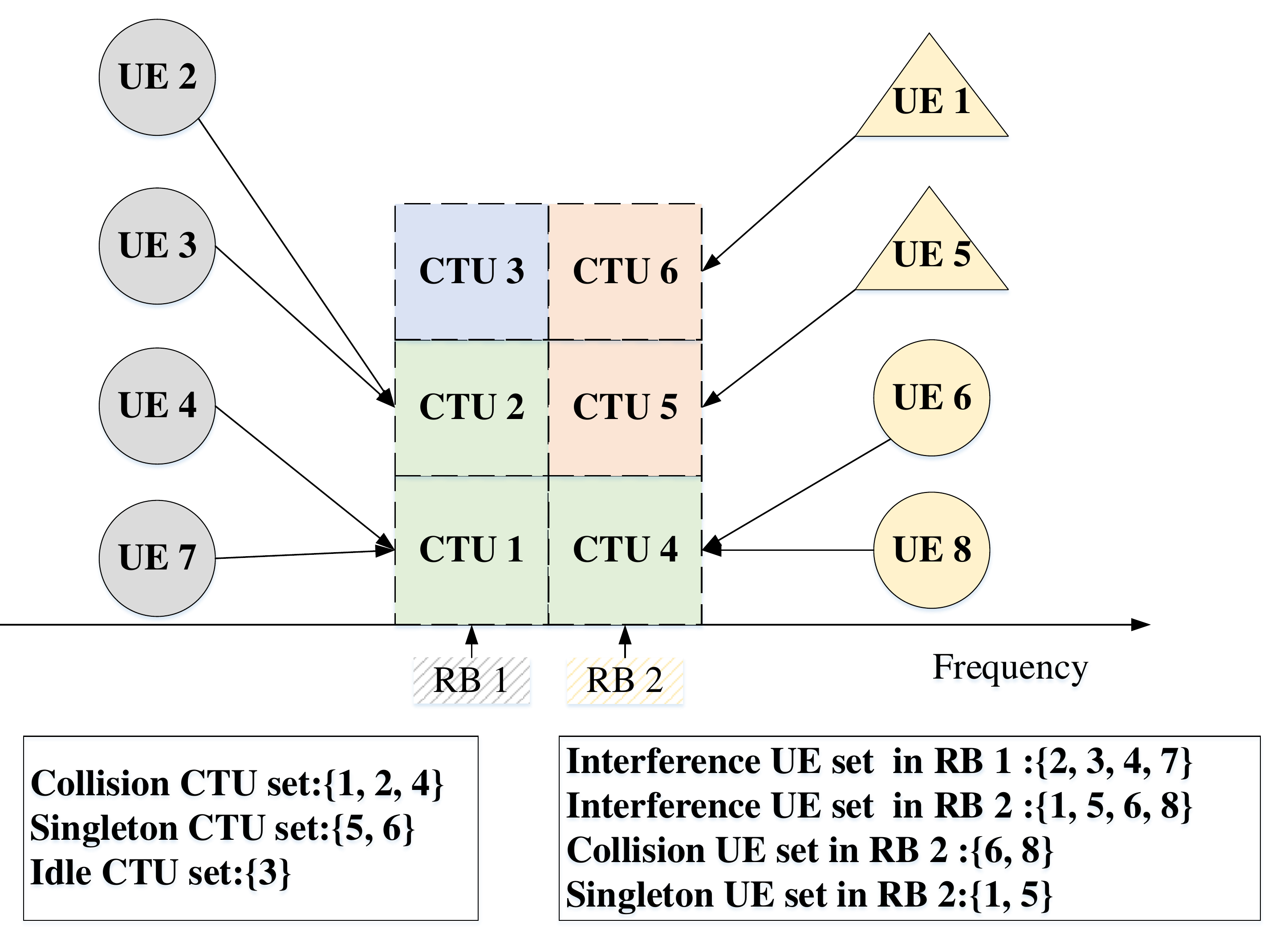}
	\caption{Detection and Decoding in a network with L=2 RBs, C = 6 CTUs and N = 8 UEs.
	}
	\label{fig:7}
\end{figure}
However, if multiple   UEs choose the same CTU, the BS cannot differentiate the  these UEs and therefore cannot decode the data.
We categorize the CTUs into three types: an \textit{idle} CTU is a CTU
which has not been chosen by any UE;  
a \textit{singleton} CTU is a CTU chosen by only one UE; and a \textit{collision} CTU is a CTU chosen by two or more UEs\cite{8533378}.
One example  is illustrated in Fig.~\ref{fig:7}.
The UE 1 and UE 5 have chosen the unique CTU 6 and CTU 5, respectively, thus, the CTU 6 and 5 are singleton CTUs. 
The CTU 3 is an idle CTU. 
The UE 4 and UE 7 have chosen the CTU 1, the UE 2 and UE 3 have chosen the CTU 2, and the UE 6 and UE 8 have chosen the CTU 4,
thus, CTU 1, 2 and 4 are  collision CTUs. 
After collision detection at the $t$th RTT, the BS observes  the set of singleton CTUs  ${\cal C}_{sc}^t$, the set of idle CTUs  ${\cal C}_{ic}^t$, and the  set of collision CTUs   ${\cal C}_{cc}^t$ as shown in orange, blue and green color, respectively, in Fig.~\ref{fig:7}.

%UEs that choose the same CTU cannot be distinguished by the BS and are said to have  CTU collision.
%In the rest of this paper, we will refer to each  PS (CTU) as a layer. 
%Thus, in Fig. 8(b), a PS collision has occurred over the sixth layer. 
%\subsubsection{Detection ofUser Activity}

%UEs that choose the same pilot sequence cannot be distinguished by the BS and are said to have pilot collision.
%The BS can distinguish between collision PSs and singleton PSs; however, for collision PSs, the BS does not know thenumber of colliding UEs and is, thus, unable to decode the collision PSs. 
% Instead, the BS treats them as interference and, thus, is only concerned with estimating their aggregate power.
%We denote ${\cal D}_s$ as the set of UEs that have chosen the singleton PSs, ${\cal L}_s$ as the set of singleton PSs, ${\cal L}_c$ as the set of collision PSs, and ${\cal L}_i$ as the set of idle PSs, respectively.

\subsubsection{
Data Decoding}
%The BS is unable to decode the collision PSs.
%Instead, the BS treats them as interference. 
%and, thus, is only concerned with estimating their aggregate power.
After detecting the UEs that have chosen the singleton CTUs (e.g., UE 1 and UE 5 shown as triangle in Fig.~\ref{fig:7}), the BS performs the successive interference cancellation (SIC) technique to decode the data of these UEs.
During the decoding, the UEs that transmit in different RBs do not interfere with each other due to the orthogonality, and  only UEs that transmit in the same RB cause interference, i.e., as shown in Fig.~\ref{fig:7}, the interference UE set in RB 1 is $\{2,3,4,7\}$ shown in color grey and the interference UE set in RB 2 is $\{1,5,6,8\}$ shown in color yellow.
In each iterative stage of SIC decoding, the CTU with the strongest received power is decoded by treating the  received powers of other CTUs over the same RB  as the interference.
Each iterative stage of SIC decoding is successful when the signal-to-interference-plus-noise ratio (SINR) in that stage is larger than the SINR threshold.
If the received signal is decoded successfully, the decoded signal is subtracted from the received signal\footnote{We assume perfect SIC  the same as \cite{8533378}, with no error propagation between iterations.}. 
%In the second stage, the second strongest received power is decoded by regarding the remaining received powers as interference.
Thus, in the $k$th repetition of the $t$th RTT,  the $s$th stage of SIC decoding is successful if the SINR is higher than a threshold $\gamma_{th}$ \cite{8533378}, i.e.,
\begin{align}\label{SIC}
{\rm SINR}^{t}_{f,s}(k) = \frac{{{P}{h_{s,k}}{r_s}^{ - \eta }}}{{\sum\limits_{m=s+1}^{ {  N}_{f,sc}^t(k)} {{P_m}} {h_{m,k}}r_m^{ - \eta } + \sum\limits_{n' \in {\cal  N}_{f,cc}^t(k)}^{} {{P_{n'}}} {h_{n',k}}r_{n'}^{ - \eta } + {\sigma ^2}}} \ge {\gamma _{th}}, 
\end{align}
where $P$ is the  transmission power, ${\cal N}_{f,sc}^t$ is the set of other devices that have chosen the singleton CTUs over the $f$th RB, ${\cal N}_{f,cc}^t$ is the set of devices that have chosen the collision CTUs over the $f$th RB, 
$\sigma^2$ is the noise power.

The SIC procedure stops when one iterative stage of the SIC fails or when there are no more signals to decode.
The SIC decoding procedure for each GF scheme is described in the following.
% The SIC decoding procedure for each GF scheme is given in Fig.~\ref{fig:8} with the details described in the following.
% \begin{figure}
% \centering
% \subfigure[K-repetition scheme]
% {\includegraphics[width=2.5in,height=3.6in]{sic_krep.pdf}}
% \subfigure[Proactive scheme]
% {\includegraphics[width=2.7in,height=4.6in]{sic_proa.pdf}}
% \caption{SIC  decoding procedure for each GF scheme.}
% \label{fig:8}
% \end{figure}

$i$) K-repetition scheme:
For the K-repetition scheme as shown in Fig. 4, the  successful decoding event  occurs at least one repetition decoding succeeds. Thus, the SIC decoding procedure follows: 
\begin{itemize}
\item Step 1: Start the $k$th repetition with the initial  $k=1$, ${\cal N}_{f,sc}^t$ and ${\cal N}_{f,cc}^t$;

\item Step 2: Decode the $s$th CTU  with the initial  $s=1$ using \eqref{SIC};
% Initialize the latency constraint $\cal T$=1 TTI, and $
% {\cal P}_{F}^{\rm Reac}[T_{\rm latency}\leqslant{1}]=0$;
\item Step 3: If the $s$th CTU is successfully decoded, put the decoded UE in set ${\cal N}_{f,sd}^t(k)$ and go to Step 4, otherwise go to Step 5;
\item Step 4: If $s \le { N}_{f,sc}^t$, do $s=s+1$, go to Step 2, otherwise go to Step 5;
\item Step 5: SIC for the $k$th repetition stops;
\item Step 6: If $k \le K_{\rm Krep}$, do $k=k+1$, go to Step 1, otherwise go to the end.
% \item Step 5: Repeat the step 3 to 4 until $m=M$ and calculate the latent access failure probability $
% {\cal P}_{\rm out}^{\rm Reac}[T_{\rm latency}\leqslant{\cal T}]$ using (\ref{krep_out}).
\end{itemize}
%Finally, the set ${\cal N}_{f,sd}^t = \bigcup\limits_{k = 1}^{{K_{Krep}}} {( {{\cal N}_{f,sd}^t(k)} )} $ is the successfully decoded UEs over the $f$th RB and ${\cal N}_{sd}^t = \bigcup\limits_{f = 1}^{{F^t}} {( {{\cal N}_{f,sd}^t} )} $ is the successfully decoded UEs over the in the $t$th RTT.

$ii$) Proactive scheme:
For the Proactive scheme as shown in Fig.~\ref{fig:5}, the successful decoding event occurs once the repetition decoding succeeds. The successfully decoded UEs will not transmit in the remaining repetitions in this RTT to reduce interference to other UEs.
It should be noted that the ACK/NACK feedback can only be received after 3TTIs, which means the ACK feedback of the $k$th successful repetition can be received by the UE in the $(k+3)$th repetition and the UE stops transmission from the $(k+4)$th repetition.
In addition, the BS does not send any  ACK/NACK  feedback to the collision UEs. The collision USs in the $k$th repetition that do not receive feedback at the pre-defined timing after the UEs sent the packet (e.g., after 3TTIs) will not transmit in the remaining repetitions to reduce interference to other UEs.
\begin{itemize}
\item Step 1: Initialize  $k=1$, ${\cal N}_{f,sc}^t$ and ${\cal N}_{f,cc}^t$. If $k< 4$, go to Step 3, otherwise go to Step 2;
\item Step 2: Update the ${\cal N}_{f,sc}^t(k)={\cal N}_{f,sc}^t(k)-{\cal N}_{f,sd}^t(k-4)$ and ${\cal N}_{f,cc}^t(k)=\varnothing$;
\item Step 3: Start the $k$th repetition with  $k$, ${\cal N}_{f,sc}^t(k)$ and ${\cal N}_{f,cc}^t(k)$;
\item Step 4: Decode the $s$th CTU  with  initial  $s=1$ using \eqref{SIC};
% Initialize the latency constraint $\cal T$=1 TTI, and $
% {\cal P}_{F}^{\rm Reac}[T_{\rm latency}\leqslant{1}]=0$;
\item Step 5: If the $s$th CTU is successfully decoded, put the decoded UE in set ${\cal N}_{f,sd}^t(k)$ and go to Step 6, otherwise go to Step 7;
\item Step 6: If $s \le { N}_{f,sc}^t$, do $s=s+1$, go to Step 4, otherwise go to Step 7;
\item Step 7: SIC for the $k$th repetition stops;
\item Step 8: If $k \le K_{\rm Pro}$, do $k=k+1$, go to Step 1, otherwise go to the end.
% \item Step 5: Repeat the step 3 to 4 until $m=M$ and calculate the latent access failure probability $
% {\cal P}_{\rm out}^{\rm Reac}[T_{\rm latency}\leqslant{\cal T}]$ using (\ref{krep_out}).
\end{itemize}
Finally, the set ${\cal N}_{f,sd}^t = \bigcup\limits_{k = 1}^{{K_{Krep}}} {( {{\cal N}_{f,sd}^t(k)} )} $ is the successfully decoded UEs over the $f$th RB and ${\cal N}_{sd}^t = \bigcup\limits_{f = 1}^{{F^t}} {( {{\cal N}_{f,sd}^t} )} $ is the successfully decoded UEs  in the $t$th RTT.

%SIC stops when a layer is decoded unsuccessfully or when the decoding stage is up to three.
%We denote ${{\cal D}_{s,suc}^s}$ as the set of UEs that have been decoded successfully.
% For SIC, the detection of collision layers can be implemented through a simple algorithm. 
% At first, the BS can  distinguish between collisionlayers and singleton layers; however, for collision layer
% assumes that all the layers are singleton layers and orders the layers according to their respective received powers in descending order. 
% In the first iteration, the BS attempts to decode the strongest layer. 
% The BS can verify that the decoded information is correct through a simple CRC-check. 
% If decoding is successful, the decoded signal is cancelled from the received signal. 
% If not, the BS assumes that this is a collision layer.
% In the second iteration, the BS attempts to decode the second strongest layer while regarding the previously undecoded layer
% as interference. 

\subsubsection{HARQ Retransmissions}
%The ACK/NACK feedback mechanism depends on not only whether data decoding is successful, but also how well the UE activity detection goes. 
We We take into account  the GF-NOMA HARQ retransmissions to achieve high reliability performance.
However, due to the latency constraint $T_{\rm cons}$, the HARQ retransmission times are limited as shown in Fig.~\ref{fig:3}. The UE determines a re-transmission or not based on the following two different scenarios.

$i$) when the UE receives an ACK from the BS, it  means that the BS successfully detected the UE (i.e., the UE choosing the singleton CTUs) and  decoded the UE's data (i.e., SIC succeeds), no further re-transmission is needed;

$ii$) when the UE receives a NACK from the BS, it means that the BS successfully detected the UE  but failed to decode the UE's data (i.e., SIC fails). Otherwise, when the UE does not receive any feedback at the pre-defined timing after the UE sent the packet (e.g., at the end of one RTT), it means the BS failed to identify the UE, the UE determines whether to retransmit or not in the next RTT based on the transmission latency check as shown in Fig.~\ref{fig:3}.

%two cases can be further considered with potential solutions with respect to HARQ feedback:

%In later case since the timing of HARQ feedback is known to the UE, even if the UE does not receive an ACK from BS at the expected timing, the UE can assume the packet is lost and a re-transmission can be send. 

% \begin{figure}
% 	\centering
% 	\includegraphics[width=5.2in,height=2.5in]{HARQ.PNG}
% 	\caption{HARQ retransmission structure.
% 	}
% 	\label{fig:9}
% \end{figure}

%In GF-NOMA, the data immediately follows the pilot, thus allowing the potential for low latency communication. In this case, however, the UE may receive either a data ACK, a pilot ACK or neither. If it receives a data ACK, it is done. If it receives a pilot ACK, it understands that its pilot  was detected but its data was not; hence, it will re-transmit the data at a known time/frequency offset. Finally, if neither data nor pilot ACK was received, it re-transmits the pilot and data. 

\subsection{Problem Formulation}
%We focus the uplink contention-based GF-NOMA procedure  for two GF schemes  over a set of preconfigured MA resources.
%Each UE has only two possible states, either\textit{inactive} or \textit{active}, while a UE with small data packets to be transmitted is in the active state. 
%A  random set of UEs are active at one given time. 
Once actived in a given RTT $t$, a UE executes the GF-NOMA  procedure, where the UE  randomly chooses one of the preconfigured $C^t$ CTUs to transmit its packets for $K_{\rm{Krep}}^t$ times or  $k_{\rm{Proa}}^t\le K_{\rm{Proa}}^t$ times under the K-repetition scheme  and the Proactive scheme, respectively. 
During this RTT, the GF-NOMA fails if: ($i$) a CTU collision occurs when two or more UEs choose the same CTU  (i.e., UE detection fails); or ($ii$) the SIC decoding fails (i.e., data decoding fails).
%or ($iii$) the transmission latency $T_{\rm late}>T_{cons}$.
Once failed, UEs decides whether to retransmit in the following RTT or not based on the transmission latency check. When $T_{\rm late}>T_{\rm cons}$, the UE fails to be served  and its packets will be dropped.
It is obvious that 1) increasing the repetition values  $K^t$ could improve the GF-NOMA success probability, but  results in an increasing latency;
2) increasing CTU numbers $C^t$ could improve the UE detection success probability, but it results in low resource utilization efficiency.
%It is necessary to optimize these parameters.

Thus, it is necessary to tackle the problem of optimizing the GF-NOMA configuration defined by parameters\footnote{According to the UE detection and data decoding procedure described in Section II.A, for the same  CTU number $C^t$, a large RB number $F^t$ leads to fewer UEs in each RB, which increases the data decoding success probability. That is to say, the larger RB number, the better. Thus, we fix the RB number $F=4$ in this work to optimize the CTU number.}  $A^t$ = $\{K^t,C^t\}$
for each RTT $t$ under both the K-repetition scheme and the Proactive scheme, where $K^t$ is the repetition value and $C^t$ is the number of CTUs.
%In this work, we focus on the most common GF schemes, which are K-repetition and Proactive schemes.A BS can adapt the parameters of these schemes in an online manner,  including $\{K_{Krep}\}$ and $\{K_{Pro}\}$
%At the beginning of each RTT r, the decision is made by the BS according to the transmission receptions 
At the beginning of each RTT $t$, the decision is made by the BS according to the transmission receptions ${U^{t'}}$ for all prior RTTs (${t'=1,...,t-1}$), consisting of the  following variables: the number of the collision CTUs ${V_{cc}^{t'}}$, the number of the idle CTUs ${V_{ic}^{t'}}$, the  number of the singleton CTUs ${V_{sc}^{t'}}$,  the number of UEs  that have been successfully detected and decoded under the latency constraint ${V_{sd}^{t'}}$, and the number of UEs  that have been successfully detected but not successfully decoded  ${V_{ud}^{t'}}$.
We denote ${{H^t}=\{O^1, O^2,..., O^{t-1}\}}$ with ${{O^{t-1}}=\{U^{t-1}, A^{t-1}\}}$ as the observation in each RTT $t$ including  histories of all such measurements and past actions.

At each RRT $t$, the BS aims at maximizing a long-term objective $R_t$ related to the average number of UEs that have successfully send data under the latency constraint ${V_{sd}^{t'}}$ with respect to the stochastic policy $\pi$ that maps the current observation history $O^t$ to the probabilities of selecting each possible parameters in $A^t$. 
%The optimization relies on the selection of parameters in $A_t$ according to the current historical observation $O_t$ with respectto the stochastic policy $\pi$. 
This optimization problem (P1) can be formulated as:
\begin{align}\label{P1}
 ({\rm P1}:)&\max\limits_{\pi ({A^t}| {{O^t}} )}\quad \sum\limits_{k = t}^\infty {{\gamma ^{k - t}}}{{\mathbb E}_\pi }[V_{sd}^k]\\
& \begin{array}{r@{\quad}r@{}l@{\quad}l}
s.t.&T_{\rm late} \le T_{\rm cons}\\
% &x_j&\geq110,  &i=1,2,3\ldots,n  \\
\end{array},
\end{align}
where $\gamma \in [0,1)$ is the discount factor for the performance accrued in the future RTTs, and $\gamma=0$  means that the agent just concerns the immediate reward.

Since the dynamics of the GF-NOMA system is Markovian over the continuous RRTs, this is a Partially Observable
Markov Decision Process (POMDP) problem which is generally intractable.
Here, the partial observation refers to that a BS can not fully know all the information of the communication environment,  including, but not limited to, the channel conditions, the UE transmission latency, the random collision process, and the traffic statistics.
%The transition between any two successive states relies on the random CTU collision process as well as the traffic generation process.
%Furthermore, the traditional optimization methods may need global information to achieve the optimal solution, which not only increases the overhead of signal transmission, but also increase the computation complexity, or even hardly deal with. 
Approximate solutions will be discussed in Section IV and V.

\section{Preliminaries and Conventional Solutions}
The optimization problem (P1) is really complicated, which cannot be easily solved via the conventional uplink resource optimization solutions,  especially the dynamic optimization taking  into account the latency constraint.
In addition, most prior works simplified the optimization without consideration of future performance\cite{7404058}.
We modify the load estimation (LE) approach given in \cite{7404058} via estimating based on the last number of the collision CTUs ${V_{cc}^{t-1}}$ and the previous numbers of idle
CTUs ${V_{ic}^{t-1}}$, ${V_{ic}^{t-2}}$, $\cdots$, ${V_{ic}^{1}}$.
To simplify, we propose a load estimation-based uplink resource configuration (LE-URC) approach to dynamically configure the CTUs number $C^t$ with the fixed repetition value $K^t$ in each RTT to maximize the successfully served UEs without latency check and SIC procedure\footnote{In the conventional solution, we ignore the SIC detection failure. That is to say, the UE is successfully transmitted if there is no CTU collision occurs. Thus, the optimization objective is $V_{sc}^t$.}  described in Section III, which is expressed as
\begin{align}\label{P2}
 ({\rm P2}:)&\max\limits_{\pi ({C^t}| {{O^t}} )}{{\mathbb E}_\pi }[V_{sc}^t],
\end{align}

% We modify the load estimation (LE) approach given in \cite{7404058} via estimating based on the last number of the collision CTUs ${V_{cc}^{t-1}}$ and the previous numbers of idle
% CTUs ${V_{ic}^{t-1}}$,${V_{ic}^{t-2}}$,$\cdots$. 

At the  RTT $t-1$ we consider that  $D_{UE}^{t - 1}=n$ UEs randomly choose one of the available $C^{t-1}$ CTUs with an equal probability $1/C^{t-1}$. The probability that no UE
chooses a CTU $c$ is
\begin{align}\label{user probability}
    {\mathbb{P}}({D_c} = 0 | {D_{UE}^{t - 1} = n}) = {( {1 - 1/{C^{t - 1}}} )^n}.
\end{align}
The expected number of idle CTUs is given by
\begin{align}\label{idle pro}
    \mathbb{E}[ {V_{ic}^{t - 1}| {D_{UE}^{t - 1} = n} } ] = \sum\limits_{c = 1}^{{C^{t - 1}}} {\mathbb{P}({D_c} = 0 | {D_{UE}^{t - 1} = n} ) = } {C^{t - 1}}{(1 - 1/{C^{t - 1}})^n}.
\end{align}
Due to that the actual number of idle CTUs $V_{ic}^{t - 1}$ can be observed at the BS, the number of active UEs in the  $(t-1)$th RTT  is estimated as
\begin{align}\label{estimate}
    {\tilde D_{UE}^{t - 1}} = {f^{ - 1}}(\mathbb{E}[ {V_{ic}^{t - 1}| {D_{UE}^{t - 1} = n} } ]) = {\log _{(1 - 1/{C^{t - 1}})}}(V_{ic}^{t - 1}/{C^{t - 1}}).
\end{align}
Next, we need to estimate the number of active UEs in the $t$th RTT $\tilde D_{UE}^t$. We use $\delta^t$ to represent the difference between the estimated numbers of UEs in the $(t-1)$th and the $t$th RTTs. That is ${\delta ^t} = \tilde D_{UE}^t - \tilde D_{UE}^{t - 1}$ for $t=1,2,\cdots$, where $\tilde D_{UE}^{0}=0$.
According to \cite{7404058}, we have ${\delta ^t}\approx {\delta ^{t-1}}$.
Therefore, the number of UEs in RTT $t$ is estimated as
\begin{align}\label{estimate_t}
    \tilde D_{UE}^t = \max \{ {2V_{cc}^{t - 1},\tilde D_{UE}^{t - 1} + {\delta ^{t-1}}} \},
\end{align}
where $2V_{cc}^{t - 1}$ represents that there are at least $2V_{cc}^{t - 1}$ number of UEs colliding in the last RTT.

Based on the estimated number of active UEs in the $t$th RTT $\tilde D_{UE}^t$, the probability that only one UE
chooses CTU $c$ (i.e., no collision occurs) is given by
\begin{align}\label{success_up}
    {\mathbb{P}}({D_c} = 1| {\tilde D_{UE}^t = n} ) = \Big( \begin{array}{l}
n\\
1
\end{array} \Big)1/{C^t}{( {1 - 1/{C^t}} )^{n - 1}}.
\end{align}
The the expected number of the
successfully served UEs in the $t$th RTT is given as
\begin{align}\label{UE_number}
    V_{suss}^t(C^t)=\mathbb{E}[ {V_{sc}^{t }| {\tilde D_{UE}^{t} = n} } ] = \sum\limits_{c = 1}^{{C^{t}}} {\mathbb{P}({D_c} = 1 | {D_{UE}^{t } = n} ) = } n{(1 - 1/{C^{t }})^{n-1}}.
\end{align}
The maximal expected number of the successfully served UEs is obtained by choosing the number of CTUs as
\begin{align}\label{max}
    {C^{t * }} = \mathop {\arg \max }\limits_{C^t \in {{\cal N}_{CTU}}} V_{suss}^t(C^t).
\end{align}

\section{Deep reinforcement learning-based GF-NOMA resource configuration}

The deep reinforcement learning (DRL) is regarded as a powerful tool to address complex dynamic control problems in POMDP.
In this section, we propose a Deep Q-network (DQN) based algorithm to tackle  the problem (P1).
The reasons in choosing DQN are that: 1) the Deep  Neural  Network  (DNN)  function approximation is able to deal with several kinds of partially observable problems \cite{sutton2018reinforcement,mnih2015human}; 2) DQN has the potential to accurately approximate the desired value function while addressing a problem with very large state spaces; 3) DQN is with high scalability, where the scale of its value function can be easily fit to a more complicated problem; 4) a variety of libraries have been established to facilitate building DNN architectures and accelerate experiments, such as TensorFlow, Pytorch, Theano, Keras, and etc..
To evaluate the capability of DQN in GF-NOMA, we first consider the dynamic configuration of repetition value $K^t$ with fixed CTU numbers $C^t$, where the DQN agent dynamically configures the $K^t$ at the beginning of each RTT for  K-repetition and  Proactive GF schemes. We then propose a cooperative multi-agent learning technique based on the DQN to optimize the configuration of both repetition value $K^t$ and CTU numbers $C^t$ simultaneously, which breaks down the selection in high-dimensional action space into multiple parallel sub-tasks.

\subsection{Deep reinforcement learning-based single-parameter configuration}
\subsubsection{Reinforcement learning framework}
\label{RL_framework}
To optimize the number of successfully served UEs under the latency constraint in GF-NOMA schemes, we consider a RL-agent deployed at the BS to interact with the environment in order to choose appropriate actions progressively leading to the optimization goal. We  define $s \in \cal S$, $a \in \cal A$, and $r \in \cal R$ as any state, action, and reward from their corresponding sets, respectively. 
The RL-agent first observes the current state $S^t$ corresponding to a set of previous observations $(O^t=\{U^{t-1},U^{t-2},...,U^{1} \})$ in order to select
an specific action $A^t\in {\cal A}(S^t)$.
Here, the action $A^t$ represents the repetition values $K^t$ in the $t$th RTT $A^t=K^t$ in this single-parameter configuration scenario and the  $S^t$ is a set of indices mapping to the current observed information $U^{t-1}=[V_{cc}^{t-1},V_{ic}^{t-1},V_{sc}^{t-1},V_{sd}^{t-1},V_{ud}^{t-1}]$.
With the knowledge of the state $S^T$, the RL-agent chooses an action $A^t$ from the set $\cal A$. 
Once an action $A^t$ is performed, the RL-agent transits to a new observed state $S^{t+1}$ and receives a corresponding reward $R^{t+1}$ as the feedback from the environment, which is designed based on the new observed state $S^{t+1}$ and guides the agent to achieve the optimization goal. 
As the optimization goal is to maximize the number of the successfully served UEs under the latency constraint, we define the reward $R^{t+1}$ as 
\begin{align}\label{reward}
R^{t+1}=V_{sd}^t,    
\end{align}
where $V_{sd}^t$ is the  observed number of successfully served UEs under the latency constraint $T_{\rm cons}$.

%When the number of the states and actions are small, RL algorithms can efficiently obtain the optimal policy.

%By using a delayed reward, the RL-agent updates its policy $\pi$ of action $A^t$. 

To select an action $A^t$ based on the current state $S^t$, a mapping policy  $\pi(a|s)$ learned from a state-action value function $Q(s, a)$ is needed to facilitate the action selection process, which indicates probability distribution of actions with given states. Accordingly, our objective is to find an optimal value function $Q^*(s, a)$ with optimal policy $\pi^*(a|s)$. At each RTT, $Q(s, a)$ is updated based on the received reward by following 
\begin{align}\label{Qtable1}
 Q(S^t,A^t)=    Q(S^t,A^t)+\lambda[R^{t+1}+\gamma \mathop {\max }\limits_{a \in {\cal A}} Q(S^{t+1},a)-Q(S^{t},A^t) ],
\end{align}
where $\lambda$ is a constant learning rate reflecting how fast the model adapting to the problem, $\gamma \in [0, 1)$ is the discount rate that determines how current rewards affect the value function updating. After enough iterations, the BS can learn the optimal policy that maximizes the long-term rewards.

\subsubsection{Deep Q-network}
When  the state and action spaces are large, the RL algorithm becomes expensive in terms of memory and computation complexity, which is difficult to converge to the optimal solution. To overcome this problem, DQN is proposed in \cite{mnih2015human}, where the Q-learning is combined with DNN to train a sufficiently accurate state-action value function for the problems with high dimensional state space. Furthermore, the DQN algorithm utilizes the experience replay technique to enhance the convergence performance of RL. When updating the DQN algorithm, mini-batch samples are selected randomly from the experience memory as the input of the neural network, which breaks down the correlation among the training samples. In addition, through averaging the selected samples, the distribution of training samples can be smoothed, which avoids the training divergence.

In DQN algorithm, the action-state value function $Q(s.a)$ is parameterized via a function $Q(s,a,\bm{\theta})$, where \bm{$\theta$} represents the weights matrix of a multiple layers DNN.
We consider the conventional fully-connected DNN, where the neurons between two adjacent layers are fully pairwise connected. The variables in the state $S^t$ is fed in to the DNN as the input; the Rectifier Linear Units (ReLUs) are adopted as intermediate hidden layers by utilizing the function $f(x)=\max\  (0,x)$; while the output layer is consisted of linear units, which are in one-to-one correspondence with all available actions in $\cal A$.

To achieve exploitation, the forward propagation of Q-function $Q(s,a,\bm{\theta})$ is performed according to the observed state $S^t$.
The online update of weights matrix \bm{$\theta$} is carried out along each training episode to avoid the complexities of eligibility traces, where a double deep Q-learning (DDQN) training principle\cite{van2015deep} is applied to reduce the overestimations of value function (i.e., sub-optimal actions obtain higher values than the optimal action).
Accordingly, learning takes place over multiple training episodes, where each episode consists of several RTT periods. In each RTT, the parameter $\bm{\theta}$ of the Q-function
approximator $Q(s,a,\bm{\theta})$ is updated using RMSProp optimizer\cite{tieleman2012lecture} as
\begin{align}\label{RMS}
  {\bm{\theta} ^{t + 1}} = {\bm{\theta} ^{t}} - {\lambda _{\mathrm{RMS}}}\nabla {L^{\mathrm{DDQN}}}({\bm{\theta} ^t}) 
\end{align}
where $\lambda _{\mathrm{RMS}} \in (0,1]$ is RMSProp learning rate, $\nabla {L^{\mathrm{DDQN}}}({\bm{\theta}^t})$ is the
gradient of the loss function ${L^{\mathrm{DDQN}}}({\bm{\theta} ^t})$ used to train the state-action value function. The gradient of the loss function is defined as
\begin{align}\label{approximator}
    \nabla {L^{\mathrm{DDQN}}}({\bm{\theta}^t}) = {{\rm E}_{{S^i},{A^i},{R^{i + 1}},{S^{i + 1}}}}[({R^{i + 1}} + \gamma \mathop {\mathop {\max }\limits_{a \in {\cal A}} \ Q({S^{i + 1}},a,{{\bar {\bm{\theta}} }^t})}  - Q({S^i},{A^i},{\bm{\theta} ^t})){\nabla _{\bm{\theta}} }Q({S^i},{A^i},{{\bm{\theta}} ^t})].
\end{align}

We consider the application of minibatch training, instead of a single sample, to update the value function $Q(s,a,\bm{\theta})$, which improves the convergent reliability of value function $Q(s,a,\bm{\theta})$. Therefore, the expectation is taken over the minibatch, which are randomly selected from previous samples $(S_i, A_i, S_{i+1}, R_{i+1})$ for $i \in \{t-M_r,...,t\}$ with $M_r$ being the replay memory size\cite{sutton2018reinforcement}. When $t-M_r$ is negative, it represents to include samples from the previous episode. Furthermore, $\bar {\bm{\theta}}^t$  is the target Q-network in DDQN
that is used to estimate the future value of the Q-function
in the update rule, and $\bar {\bm{\theta}}^t$ is periodically copied
from the current value $ {\bm{\theta}}^t$ and kept unchanged for several episodes.

Through calculating the expectation of the selected
previous samples in minibatch and updating the ${\bm{\theta} ^{t}}$ by \eqref{RMS}, the DQN value function $Q(s,a,\bm{\theta})$ can be obtained. The detailed DQN algorithm is presented in \textbf{Algorithm 1.}

\begin{figure}[htbp!]
        \label{DQN}
        \renewcommand{\algorithmicrequire}{\textbf{Input:}}
        \renewcommand{\algorithmicensure}{\textbf{Output:}}
       % \removelatexerror
        \begin{algorithm}[H]
            \caption{DQN Based GF-NOMA Uplink Resource Configuration}%算法名字
            \LinesNumbered %要求显示行号
            \KwIn{The set of repetition values in each RTT  $K$ and Operation Iteration I.}%输入参数
           % \KwOut{output result}%输出
            Algorithm hyperparameters: learning rate $\lambda_{RMS} \in (0, 1])$, discount rate $\gamma \in [0, 1) $,  $\epsilon$-greedy rate $\epsilon \in (0, 1]$,  target network update frequency $K$;
            
            Initialization of replay memory $M$ to capacity $D$, the state-action value function $Q(S,A,{\bm\theta})$,
the parameters of primary Q-network $\bm\theta$, and the target Q-network $\bm{\bar \theta}$;
            
            \For{Iteration $\leftarrow$ 1 to I}{
                Initialization of $S^1$ by executing a random action $A^0$ and bursty traffic arrival rate $\mu^0=0$;
                
                \For {t $\leftarrow$ 1 to T}{
                Update $\mu^0$ using Eq. (2);
                
                \textbf{if} {$p_\epsilon <\epsilon$} \textbf{Then} select a random action $A^t$ from $\cal A$;
                
                \textbf{else} select
                ${A^t} = \mathop {\arg \max }\limits_{a \in A} Q({S^t},a,\bm\theta )$.
                The BS broadcasts $K(A^t)$ and backlogged UEs attempt communication in the $t$th RTT;
                
                The BS observes state $S^{t+1}$, and calculate the related reward $R^{t+1}$ using Eq. (15);
                
                Store transition $(S^t, A^t, R^{t+1}, S^{t+1})$ in replay memory $M$;
                
                Sample random minibatch of transitions $(S^t, A^t, R^{t+1}, S^{t+1})$ from replay memory $M$
                
                Perform a gradient descent step and update parameters $\bm\theta$ for $Q(s,a,{\bm\theta})$ using Eq. (18);
                
                Update the parameter $\bm{\bar\theta}=\bm{\theta}$ of the target Q-network every $J$ steps.
                }
            }
        \end{algorithm}
    \end{figure}

\begin{figure}[htbp!]
	\subfigure
	{\includegraphics[width=5.9in,height=3.4in]{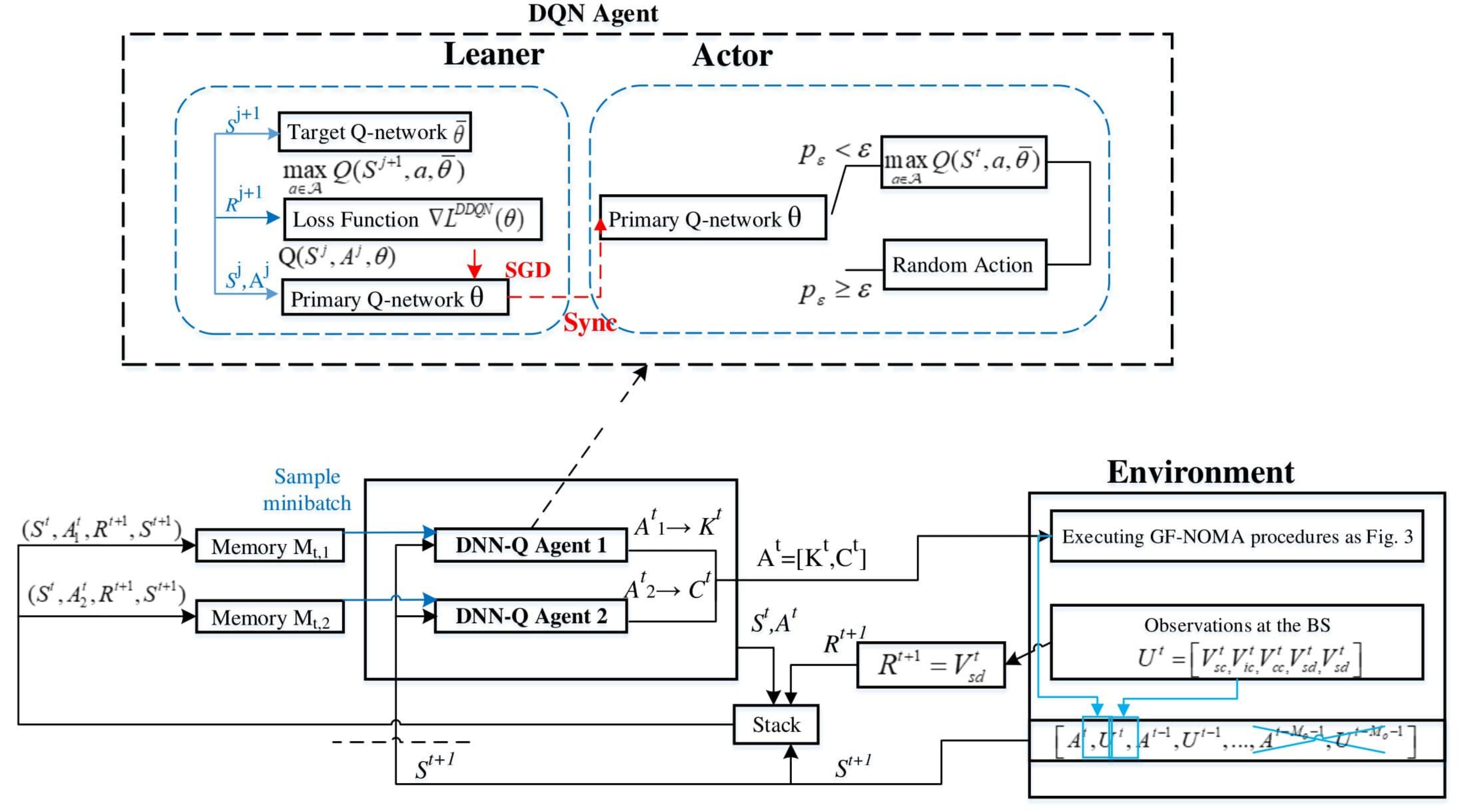}}
	\caption{The CMA-DQN agents and environment interaction in the POMDP.}
	\label{fig:9}
\end{figure}
\subsection{Cooperative Multi-Agent Learning-based multi-parameter optimization}
In practice, not only the repetition values but also the CTU numbers, influence reliability-latency performance in GF-NOMA. Fixed CTU numbers cannot adapt to the dynamics of the random traffic, which may violate the stringent latency requirement or lead to low resource efficiency. Thus, we study the problem (P1) of jointly optimizing
the resource configuration with parameters $A^t=\{K^t, C^t\}$ to improve the network performance.
The learning algorithm  provided in Sec. V.A is model-free, and thus the learning structure
can be extended in this multi-parameter scenario. 

Due to the high capability of DQN to handle problems with massive state spaces, we consider to improve
the state spaces with more observed information to support
the optimization of RL-agent. Therefore, we define the current state $S^t$, to include information about the last $M_o$ RTTs $(U^{t-1}, U^{t-2}, U^{t-3}, ... , U^{t-M_o} )$, which enables the RL-agent to estimate the trend of traffic. Similar to the state spaces, the available action spaces
also exponentially increases with the increment of the
adjustable parameter configurations in GF-NOMA. The total number of available actions corresponds to the possible combinations of all parameter configurations.

Although the GF-NOMA configuration is managed
by a central BS, breaking down the control of multiple parameters as multiple sub-tasks is sufficient to deal with the problems with unsolvable action space, which are cooperatively handled by independent Q-agents. 
As shown in Fig.~\ref{fig:9}, we consider multiple DQN agents that are centralized at the BS following the same structure of value function approximator as Sec. V.A. Each DQN agent controls their own action variable, namely $K^t$ or $C^t$, and receives a common reward to guarantee the objective in P1 cooperatively.

However, the common reward design also poses challenge on the evaluation of each action, because the individual effect of specific action is deeply hidden in the effects of the actions taken by all other DQN agents. For instance, a positive
action taken by a agent can receive a misleading low reward due to other
DQN agents’ negative actions. Fortunately, in GF-NOMA scenario,
all DQN agents are centralized at the BS and share full information among each other.
Accordingly, we include the action selection histories of each
DQN agent as part of state function, and hence, the agents are able to
learn the relationship between the common reward and different combinations of
actions. To do so, we define state variable $S^t$ as
\begin{align}\label{Statevariable}
  {{S ^{t}}} = [A^{t-1}, U^{t-1}, A^{t-2}, U^{t-2}, ..., A^{t-M_{o}}, U^{t-M_{o}}],
\end{align}
where $M_{o}$ is the number of stored observations, $A^{t-1}$ is the
set of selected action of each DQN agent in the $(t-1)$th
TTI corresponding to $K^{t-1}$, and $C^{t-1}$, and $U^{t-1}$ is the set of observed transmission receptions.

In each RTT, the $k$th agent update the parameters ${\bm\theta}_k$ of the  value function $Q(s,a_{k},\bm{\theta}_{k})$ using RMSProp optimizer following Eq. \eqref{RMS}. The learning algorithm can be implemented following \textbf{Algorithm 1}. Different from the GF NOMA single-parameter configuration scenario in Section III.A, it is required to initialize two primary networks ${\bm\theta}_k$, target networks ${\bar{{\bm\theta}}_k}$ and the replay memories $M_k$ for each DQN agent. 
In step 10 of \textbf{Algorithm 1}, each agent stores their own current transactions in memory separately. In step 11 and 12 of \textbf{Algorithm 1}, the minibatch of transaction
should separately be sampled from individual memory to train the corresponding DQN agent.

\section{Simulation Results}
In this section, we examine the effectiveness of our proposed GF-NOMA schemes with DQN algorithm
via simulation and  compare the results
with LE-URC.
We adopt the standard network parameters listed in Table I following\cite{2Tel2018}, and
hyperparameters for the DQN learning algorithm are listed in Table II.
\begin{table}[htbp!]
	\centering
	\caption{Simulation Parameters}
	{\renewcommand{\arraystretch}{1.1}
		%\renewcommand{\tabcolsep}{0.15cm}
	%	\begin{tabular}{|c|c|c|c|}
	{
		\begin{tabular}{|p{5cm}|p{2cm}|p{5cm}|p{2cm}|}
			\hline
		Parameters  & Value  & Parameters & Value \\ \hline
		Path-loss exponent $\eta$ & 4 &
		Noise power $\sigma^2$ & -132 dBm \\ 
		\hline
		Transmission power $P$ & 23 dBm   &
		The received SINR threshold $\gamma_{th}$ & -10 dB  \\ \hline
		Duration of traffic $T$  &  2 s & The set of the repetition value & $\{1, 2, 4, 6, 8\}$\\ 
		\hline
		The set of the CTU number &  $\{12, 24, 36, 48\}$ & 
		Latency constraint &  2 ms and 8 ms \\ 
		\hline
		Bursty traffic parameter Beta($\alpha$,$\beta$)&  (2, 4)  & The number of bursty UEs $N$  &  20000 \\ 
		 \hline
		 Cell radius &  10 km & 
		 Duration of one TTI & 0.125 ms \\ 
		   \hline
		\end{tabular}
		}
	}
	\label{table_accord}
\end{table}
\begin{table}[htbp!]
	\centering
	\caption{Learning Hyperparameters}
	{\renewcommand{\arraystretch}{1.1}
		%\renewcommand{\tabcolsep}{0.15cm}
	%	\begin{tabular}{|c|c|c|c|}
	{
		\begin{tabular}{|p{5cm}|p{2cm}|p{5cm}|p{2cm}|}
			\hline
		Hyperparameters  & Value  & Hyperparameters & Value \\ \hline
		Learning rate  $\lambda_{RMS}$   &0.0001 &
		Minimum exploration rate $\epsilon$ & 0.1 \\ 
		\hline
	   Discount rate  $\gamma$ & 0.5   &
		Minibatch size & 32  \\ \hline
		Replay Memory  &  10000 & Target Q-network update frequency & 1000\\ 
		\hline
		\end{tabular}
		}
	}
	\label{table_accord}
\end{table}
All testing performance results are obtained by averaging over
1000 episodes.
The BS is located at the center of a circular area with a 10 km radius, and the UEs are randomly located within the cell. 
Unless otherwise stated, we consider the number of bursty UEs to be  $N=20000$. The DQN is set with two hidden layers, each with 128 ReLU units. 
In the following, we present our simulation
results of the single-repetition configuration and the
multi-parameter configuration in Section V-A and
Section V-B, respectively.
The single-repetition configuration is optimized under the latency constraint ${T_{\rm cons}}=2$ ms and the multi-parameter configuration is optimized under the latency constraint ${T_{\rm cons}}=8$ ms, respectively.

\subsection{Single-repetition configuration}
In the single-repetition configuration scenario,
we set the number of CTU as $C$ = 48.
Throughout epoch, each UE has a periodical  a bursty traffic profile (i.e., the time limited Beta profile defined in \eqref{beta} with parameters (2, 4) that has a peak around the
4000th TTI.

\begin{figure}[htbp!]
	\centering
	\includegraphics[width=2.8in,height=2.1in]{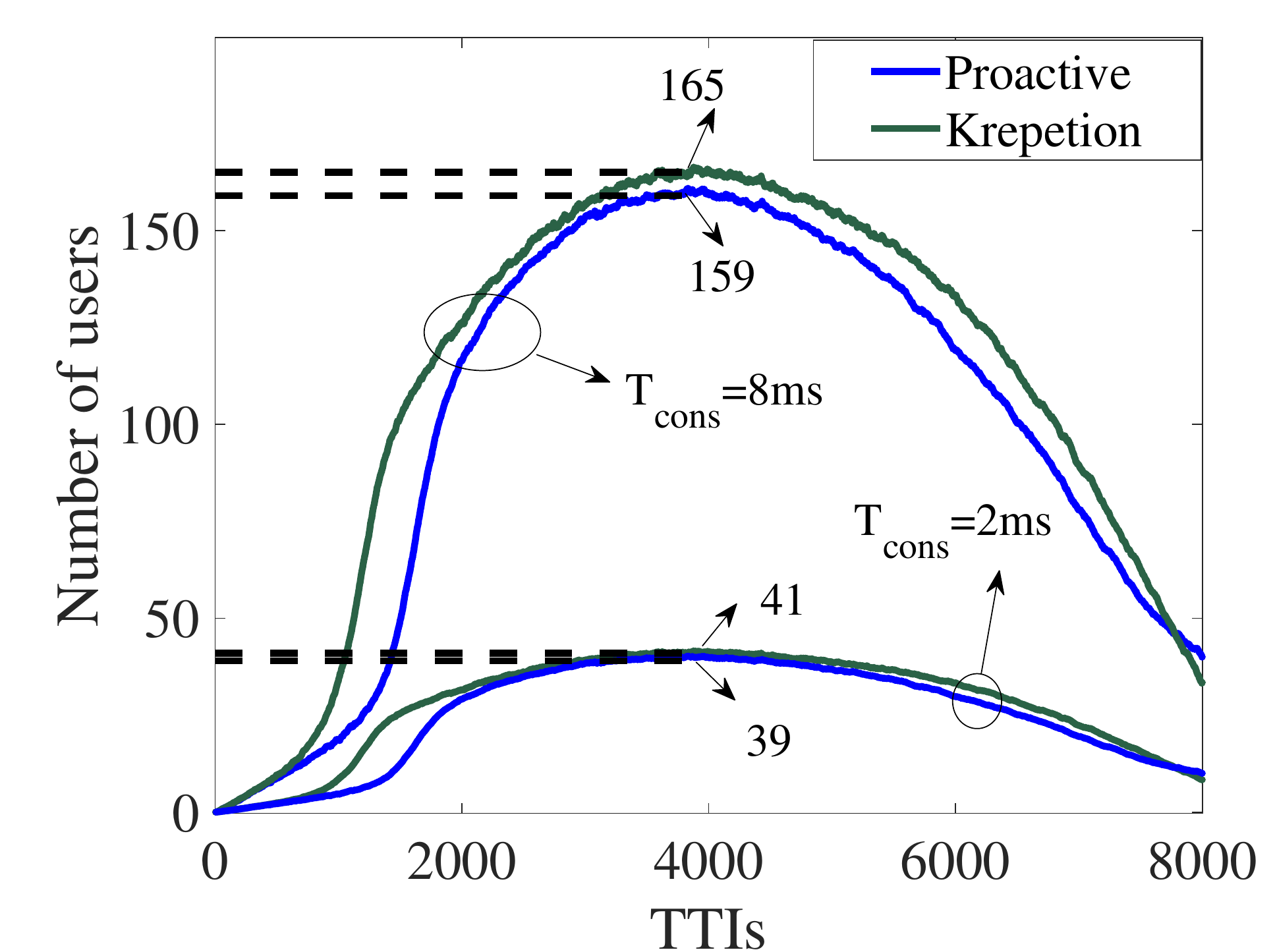}
	\caption{Backlog traffic in each TTI under latency constraint $T_{\rm cons}=2$ ms and $T_{\rm cons}=8$ ms.
	}
	\label{fig:10}
\end{figure}
Fig.~\ref{fig:10} plots the backlog traffic in each TTI under latency constraint $T_{\rm cons}=2$ ms and $T_{\rm cons}=8$ ms, respectively. 
It should be noted that the the backlog traffic in each TTI does not only include the newly generated traffic, but also the retransmission traffic, due to the fact that the UEs are allowed to retransmit in the next RTT under the latency constraint.
The results have shown that when the latency constraint increases, the backlog traffic in each TTI increase as the  retransmission traffic increases.
The backlog traffic in each TTI for the Proactive scheme is smaller than that of the K-repetition scheme due to the efficiency of the Proactive scheme, which has been analyzed in details in the following.

\begin{figure}[htbp!]
\centering
\subfigure[K-repetition scheme]
{\includegraphics[width=2.7in,height=2.1in]{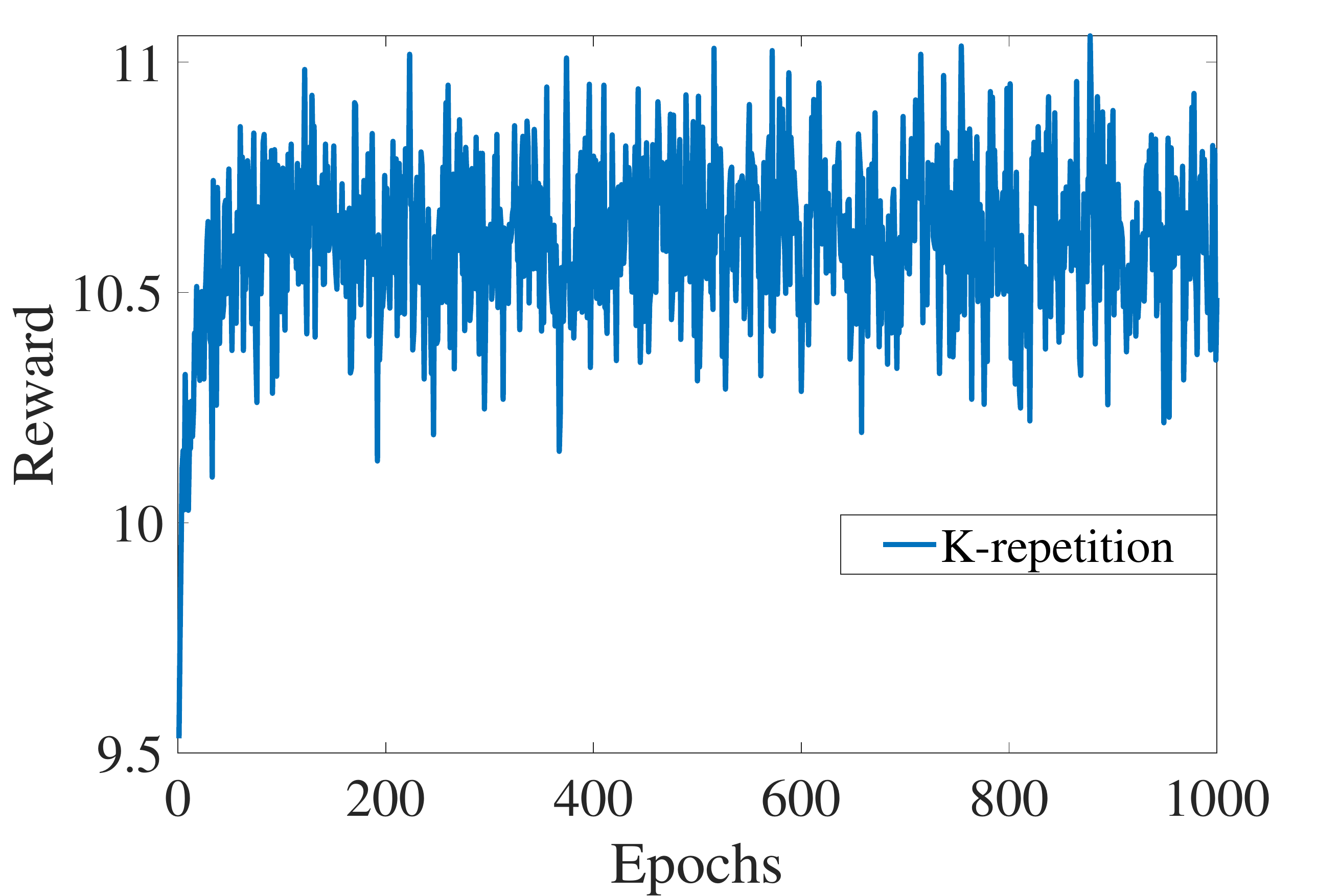}}
\subfigure[Proactive scheme]
{\includegraphics[width=2.7in,height=2.1in]{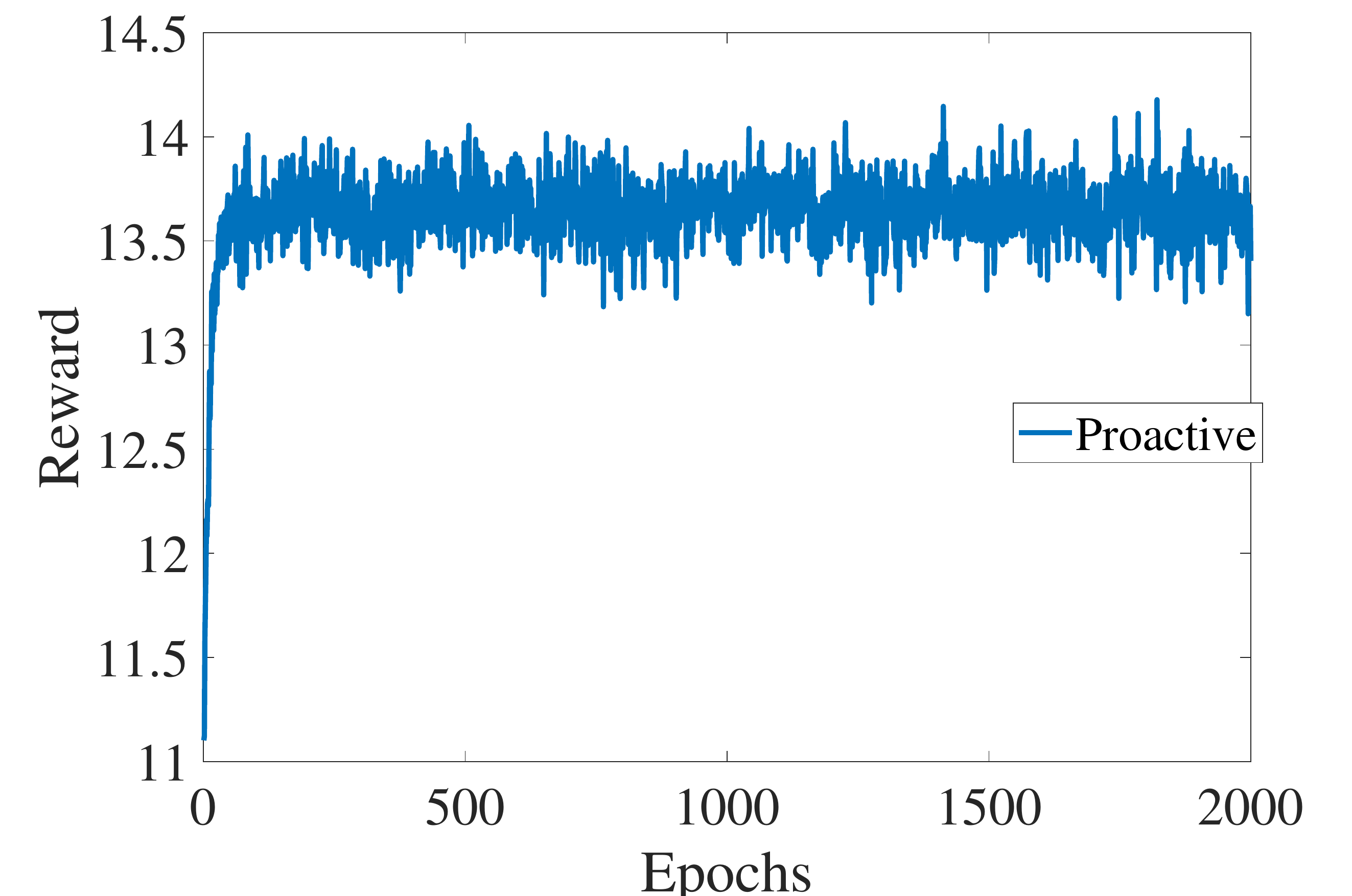}}
\caption{Average received reward for each GF scheme.}
\label{fig:12}
\end{figure}

Fig.~\ref{fig:12}  plots the average received reward for the  K-repetition scheme and the Proactive scheme, respectively. It can be seen that the average rewards of both K-repetition and proactive schemes converge to the optimal value after training. We can also observe that the average received reward of proactive scheme in Fig.~\ref{fig:12} (b) is higher than that of the K-repetition scheme in Fig.~\ref{fig:12} (a). This is because the proactive scheme can terminate the repetition earlier and start new packet transmission with timely ACK feedback, which is able to deal with the traffic more effectively.

\begin{figure}[htbp!]
\centering
{\subfigure[K-repetition scheme]
{\includegraphics[width=3.0in,height=2.1in]{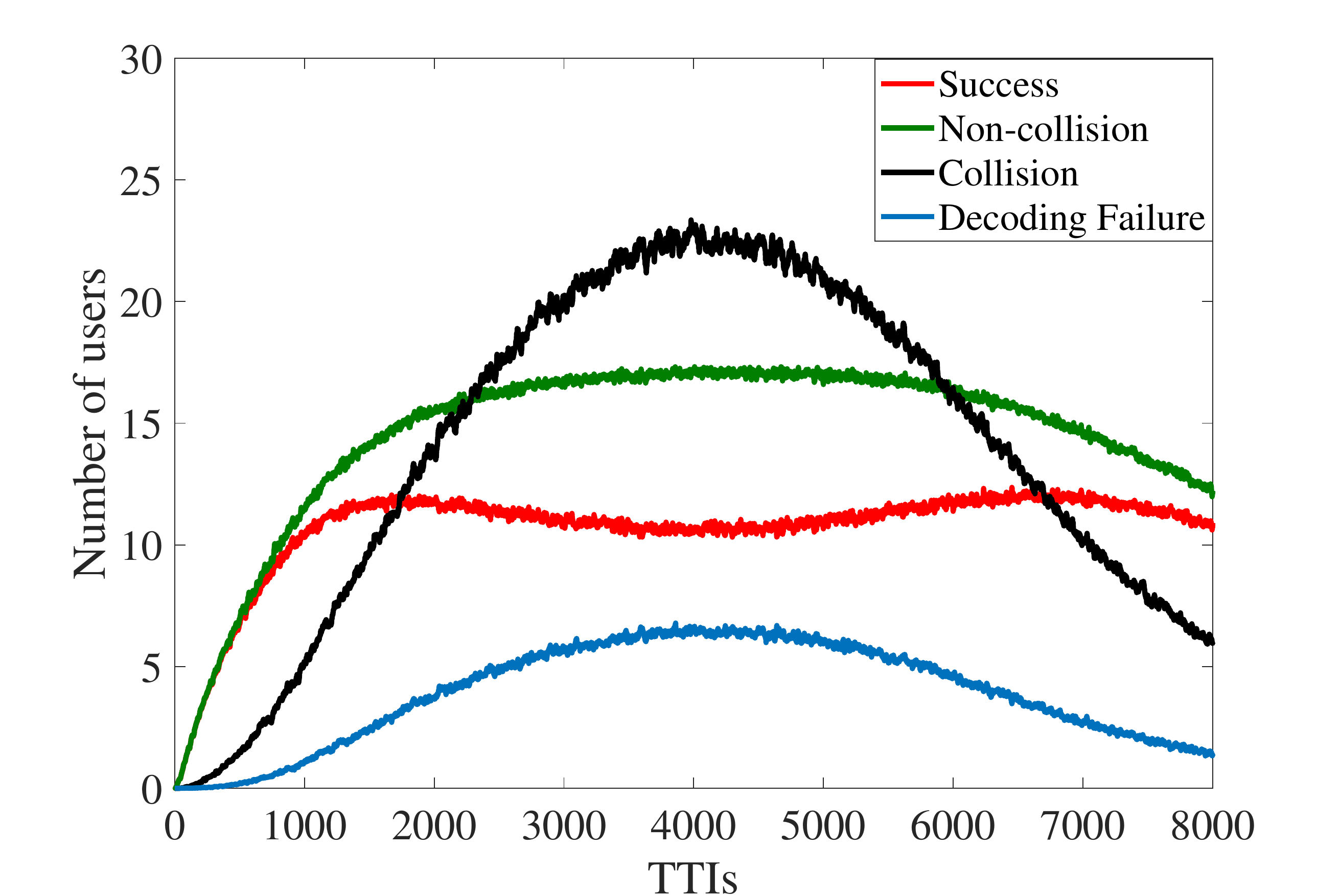}}
\subfigure[Proactive scheme]
{\includegraphics[width=3.0in,height=2.1in]{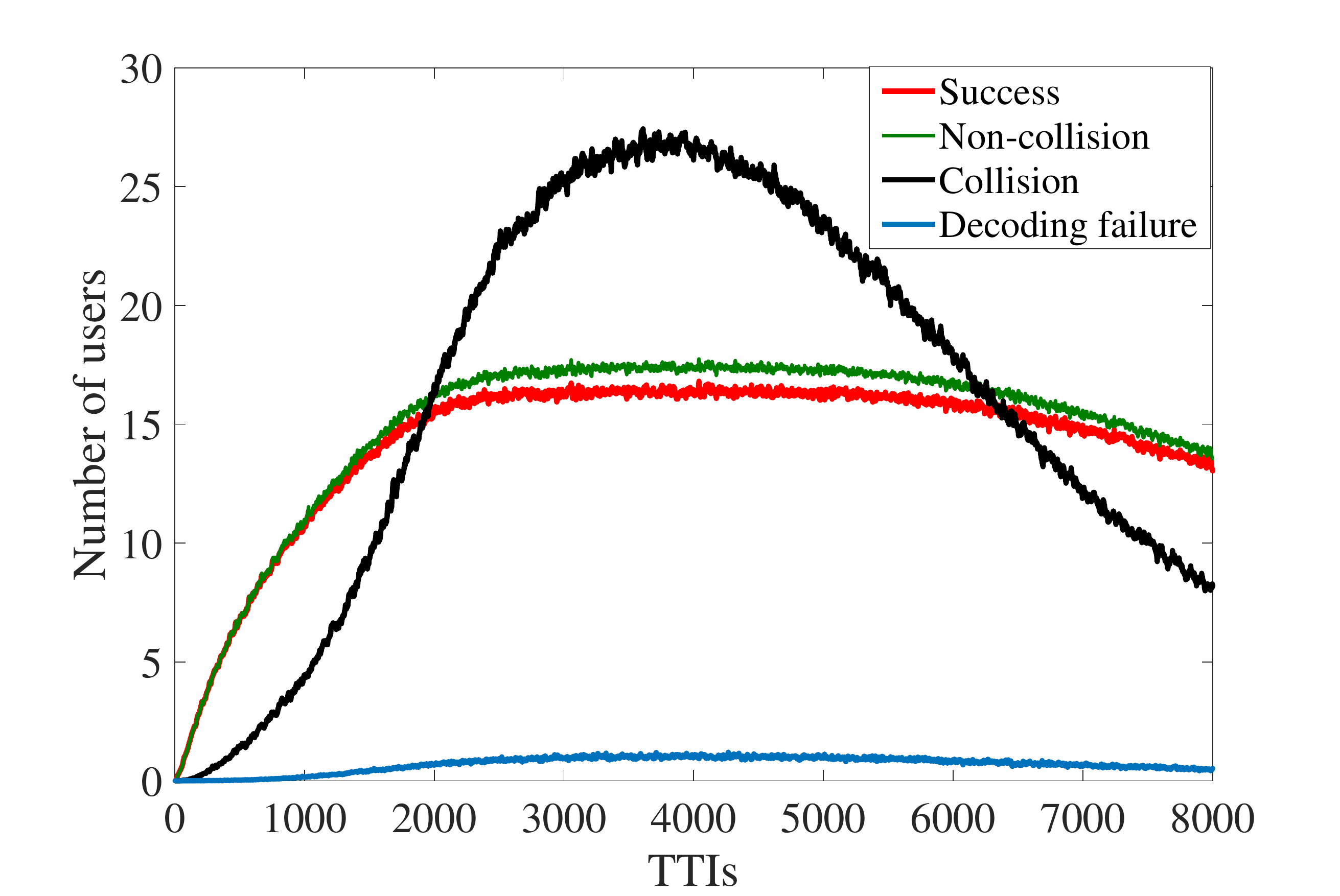}}}
\caption{The transmission results for each GF scheme.}
\label{fig:11}
\end{figure}

Fig.~\ref{fig:11}  plots the number of the successfully served UEs, the non-collision UEs, the collision UEs, and the decoding failure UEs for the K-repetition scheme and the Proactive scheme  respectively, under latency constraint $T_{\rm cons}=2$ ms.
It is shown that the number of successfully served UEs achieved under latency constraint for the Proactive scheme is almost up to 1.5 times more than that for the K-repetition scheme.
This is due to that the UEs in the Proactive scheme can terminate their repetitions earlier to reduce the interference to other UEs, which leads to an increase in the number of successfully decoding UEs.
In both Fig.~\ref{fig:11} (a) and Fig.~\ref{fig:11} (b), the number of collision UEs has a peak at around the
4000th TTI with the peak traffic at this time as shown in Fig.~\ref{fig:10}.
Due to the fact that only the non-collision UEs can be detected by the UE to decoding their data, the number of successful UEs depends on the number of the non-collision UEs.
In addition, the number of failure decoding UEs reaches a peak due to the peak traffic at the 4000th TTI, which leads to the decrease in the number of successful UEs at that time.

%It should be noted that the number of successful UEs does not only affected by the newly generated traffic but also the retransmission traffic, due to the fact that the UEs are allowed to retransmit in the next RTT under the latency constraint

\subsection{Multi-parameter configuration including the repetition values and the CTU number}
\begin{figure}[htbp!]
\centering
\subfigure[K-repetition scheme]
{\includegraphics[width=2.8in,height=2.1in]{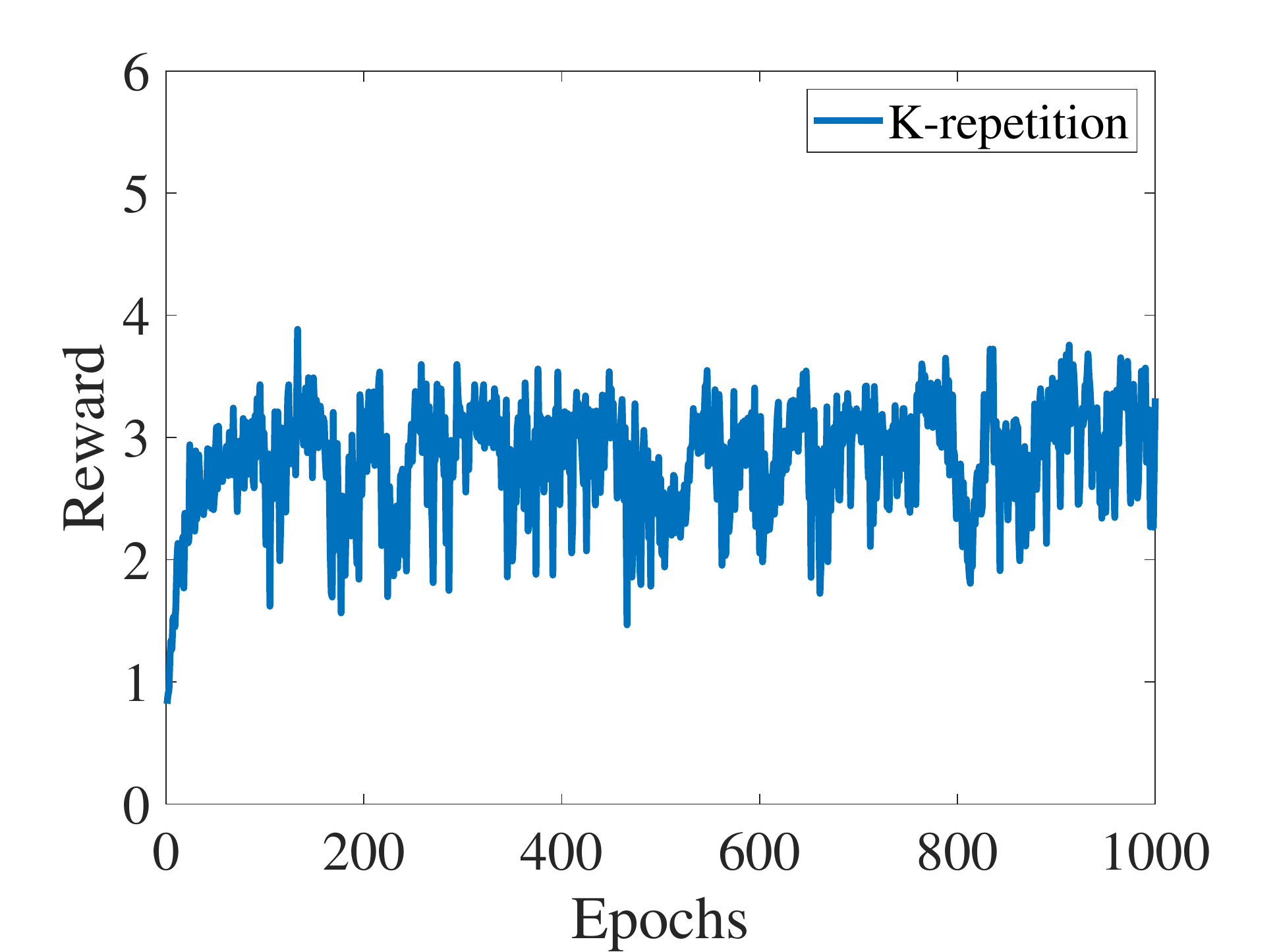}}
\subfigure[Proactive scheme]
{\includegraphics[width=2.8in,height=2.1in]{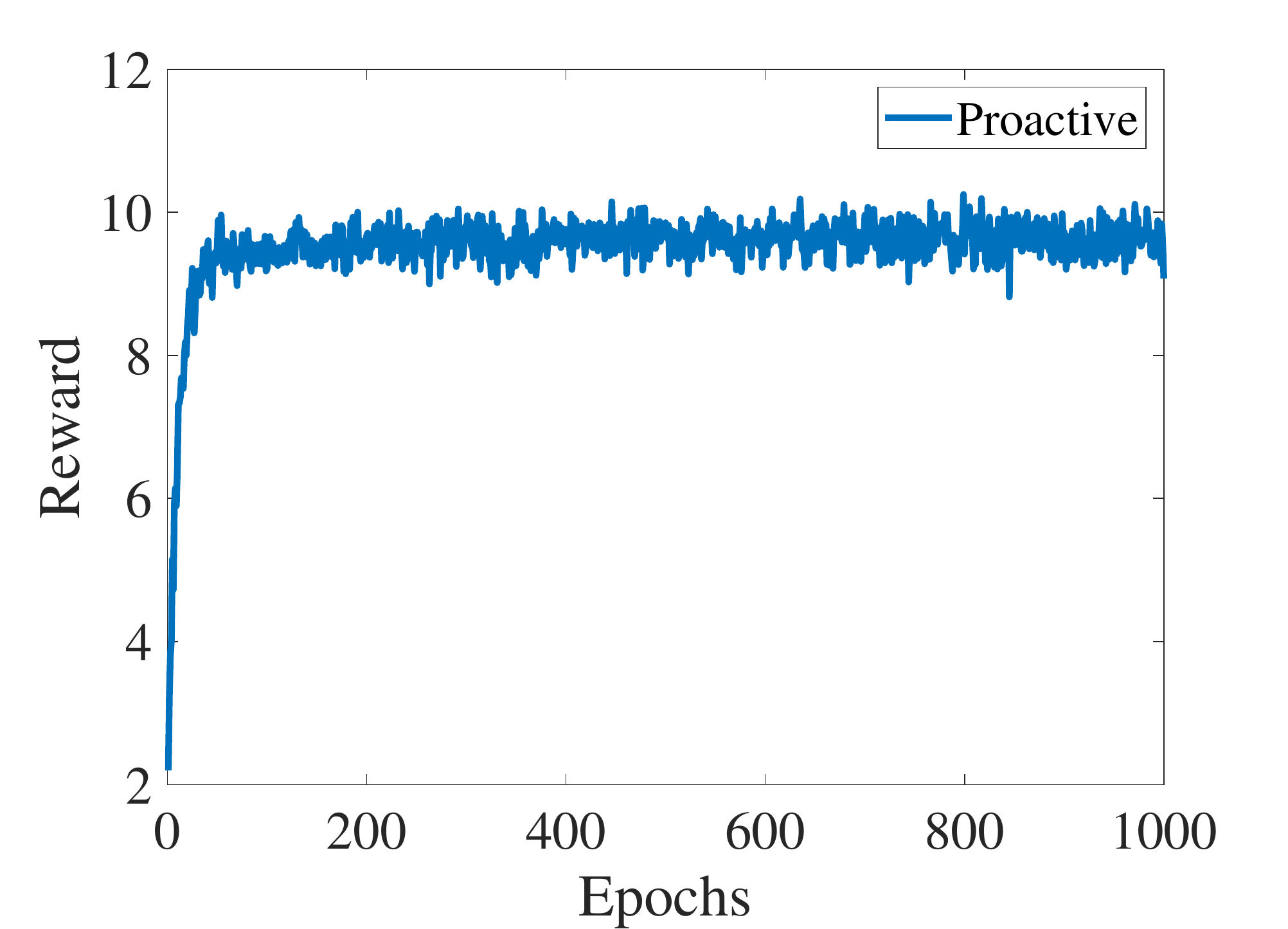}}
\caption{Average received reward for each GF scheme with multi-parameter configuration.}
\label{fig:14}
\end{figure}
Fig.~\ref{fig:14} plots the average received reward for the K-repetition scheme and the Proactive scheme with multi-parameter configuration, including the repetition values and the CTU number, respectively. It can be seen that both the average received rewards of the K-repetition and the Proactive scheme converge to a optimal value after training.
Compared to Fig.~\ref{fig:12} under latency constraint $T_{\rm cons}=$2 ms, we observe that the average received reward of two schemes decrease significantly. This is because the larger latency constraint $T_{\rm cons}=8$ ms leads to larger retransmission packets, which results in serious traffic congestion. 
It is noted that the performance degradation of K-repetition scheme is much larger than that of Proactive scheme, 
which shows the potential of the Proactive scheme in heavy traffic and long latency constraint situation due to timely
termination.

\begin{figure}[htbp!]
\centering
\subfigure[K-repetition scheme]
{\includegraphics[width=2.8in,height=2.1in]{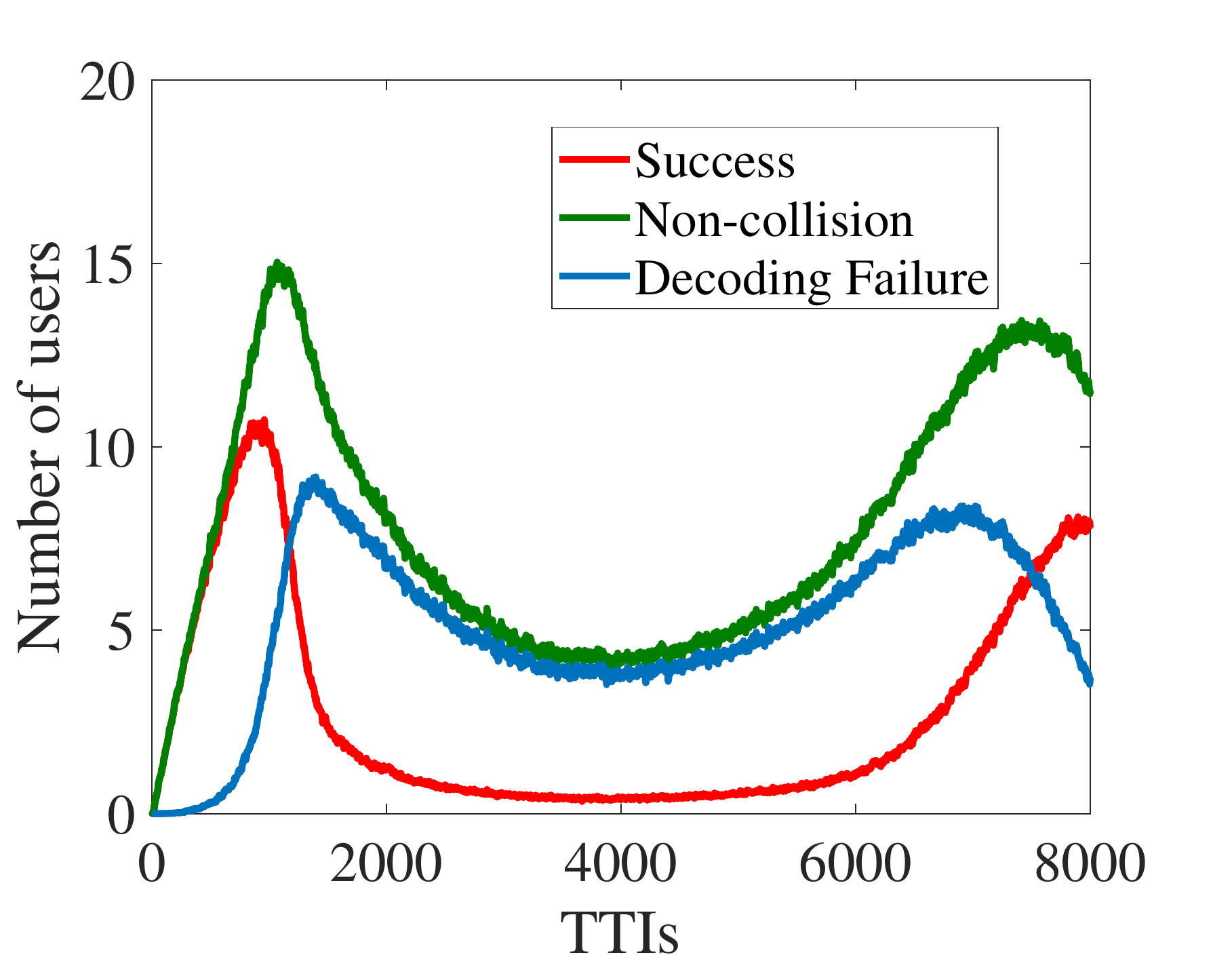}}
\subfigure[Proactive scheme]
{\includegraphics[width=2.8in,height=2.1in]{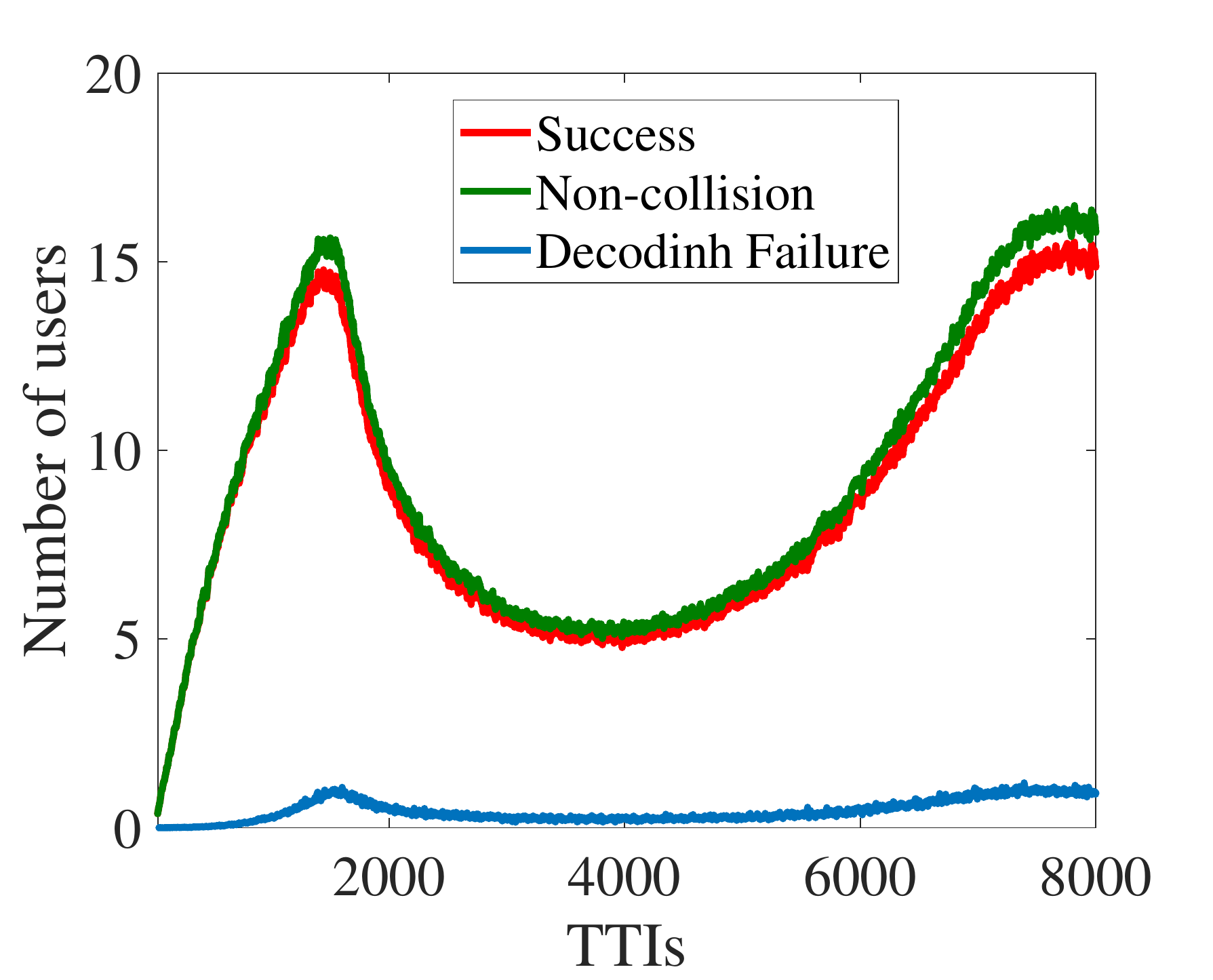}}
\caption{The transmission results for each GF scheme.}
\label{fig:13}
\end{figure}

Fig.~\ref{fig:13} plots the 
number of successful UEs, non-collision  UEs, and decoding failure
 UEs for K-repetition scheme and Proactive scheme with multi-parameter configuration including the repetition values and the CTU number under latency constraint $T_{\rm cons}=8$ ms, respectively. 
We have observed that the number of non-collision transmission UEs of both scheme is similar. However, the number of decoding failure  UEs of  the K-repetition scheme is almost up to five times more than that of the Proactive scheme at the peak traffic, due to the interference caused by  multiple repetitions from collision UEs as Fig.~\ref{fig:15}.
It is also noted that in both schemes, there is lower number of success UEs in high traffic, especially in peak traffic at round 4000th TTI.

% \begin{figure}[htbp!]
% \centering
% \subfigure
% {\includegraphics[width=2.6in,height=2.3in]{compaer.eps}}
% % \subfigure[Number of CTU]
% % {\includegraphics[width=2.6in,height=2.3in]{preamble.eps}}
% \caption{The average number of success UEs for each scheme with learning framework and fixed parameters.}
% \label{fig:16}
% \end{figure}
\begin{figure}[htbp!]
\centering
\subfigure[K-repetition scheme]
{\includegraphics[width=2.8in,height=2.1in]{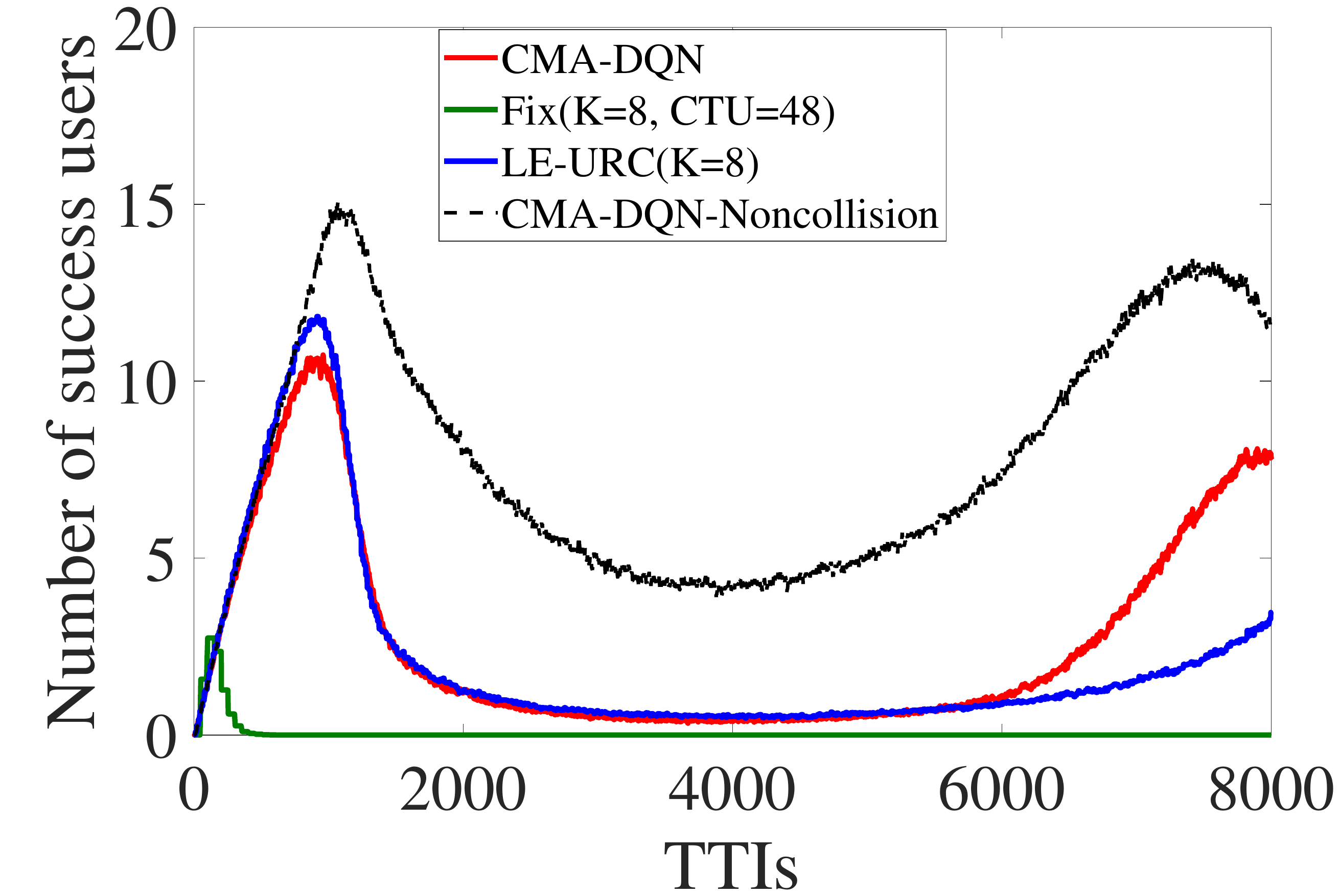}}
\subfigure[Proactive scheme]
{\includegraphics[width=2.8in,height=2.1in]{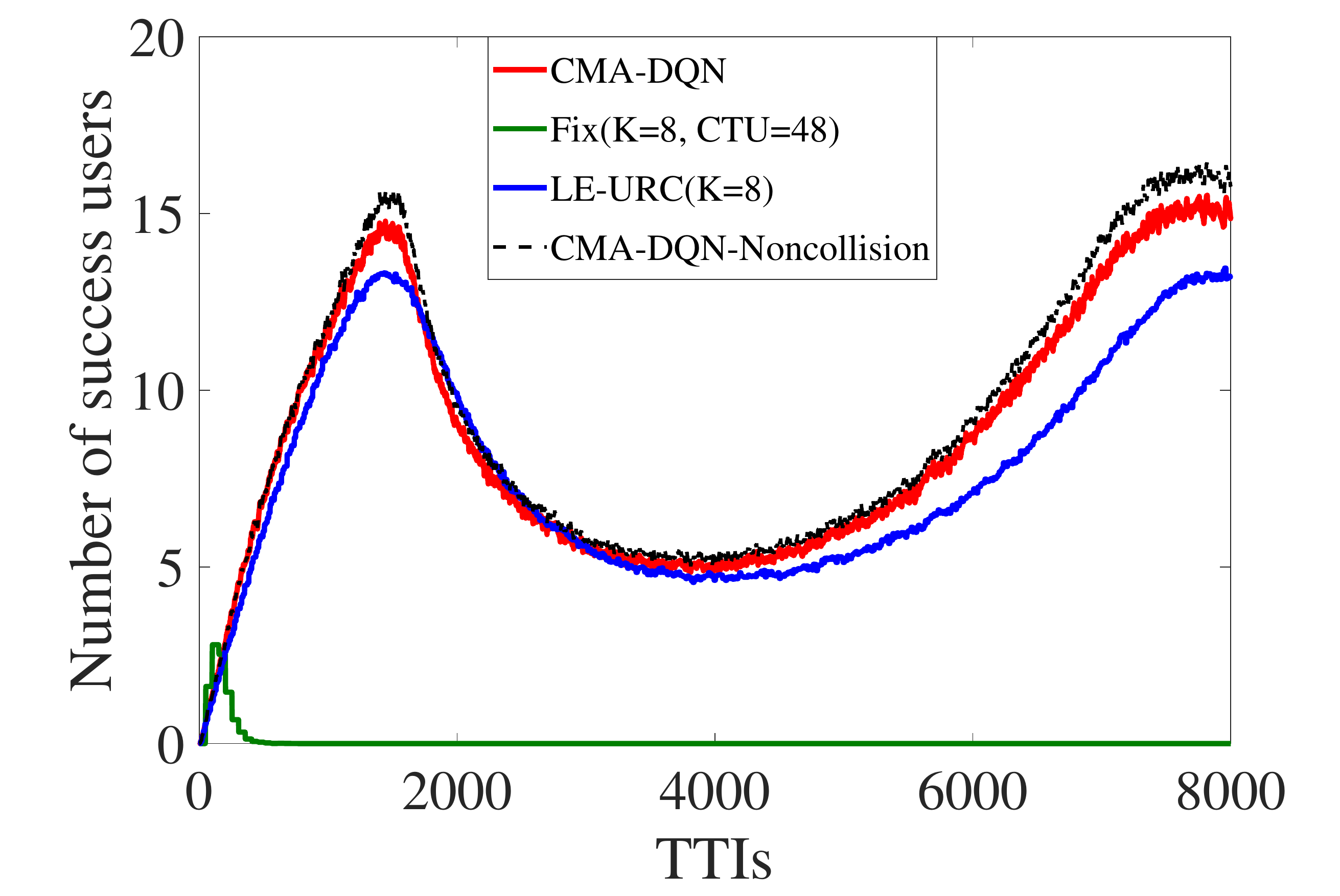}}
\caption{Average number of successfully served UEs for each GF scheme}
\label{fig:16}
\end{figure}
Fig.~\ref{fig:16} plots the average number of successful UEs for the K-repetition scheme and the Proactive scheme by comparing the learning framework with fixed parameters, and with the LE-URC approach, respectively.
Here, we set the fixed repetition value $K=8$ and the CTU number $C=48$. 
Our results  shown that the number of successfully served UEs under the same latency constraint in our proposed learning framework is up to ten times for the K-repetition scheme, and two times for the Proactive scheme, more than that with fixed repetition values and CTU numbers, respectively.
In addition, since the LE-URC approach does not aware of  the latency constraint and SIC procedure, the results are large at first, but still smaller than the number of non-collision UEs of CMA-DQN.
However, with increasing TTIs (above 6000), the cumulated traffic increases due to unsuccessful transmissions and retransmissions,  the LE-URC method becomes worse  and achieve lower number of successful UEs than that of CMA-DQN due to its  ignorance in latency constraint during its optimization for one time instance. The superior performance of CMA-DQN in heavy traffic scenario also demonstrate its capability in dynamically configure lower lower repetition
values and CTU numbers to alleviate the traffic congestion to obtain a long-term reward.

\begin{figure}[htbp!]
\centering
\subfigure[K-repetition scheme]
{\includegraphics[width=2.6in,height=2.3in]{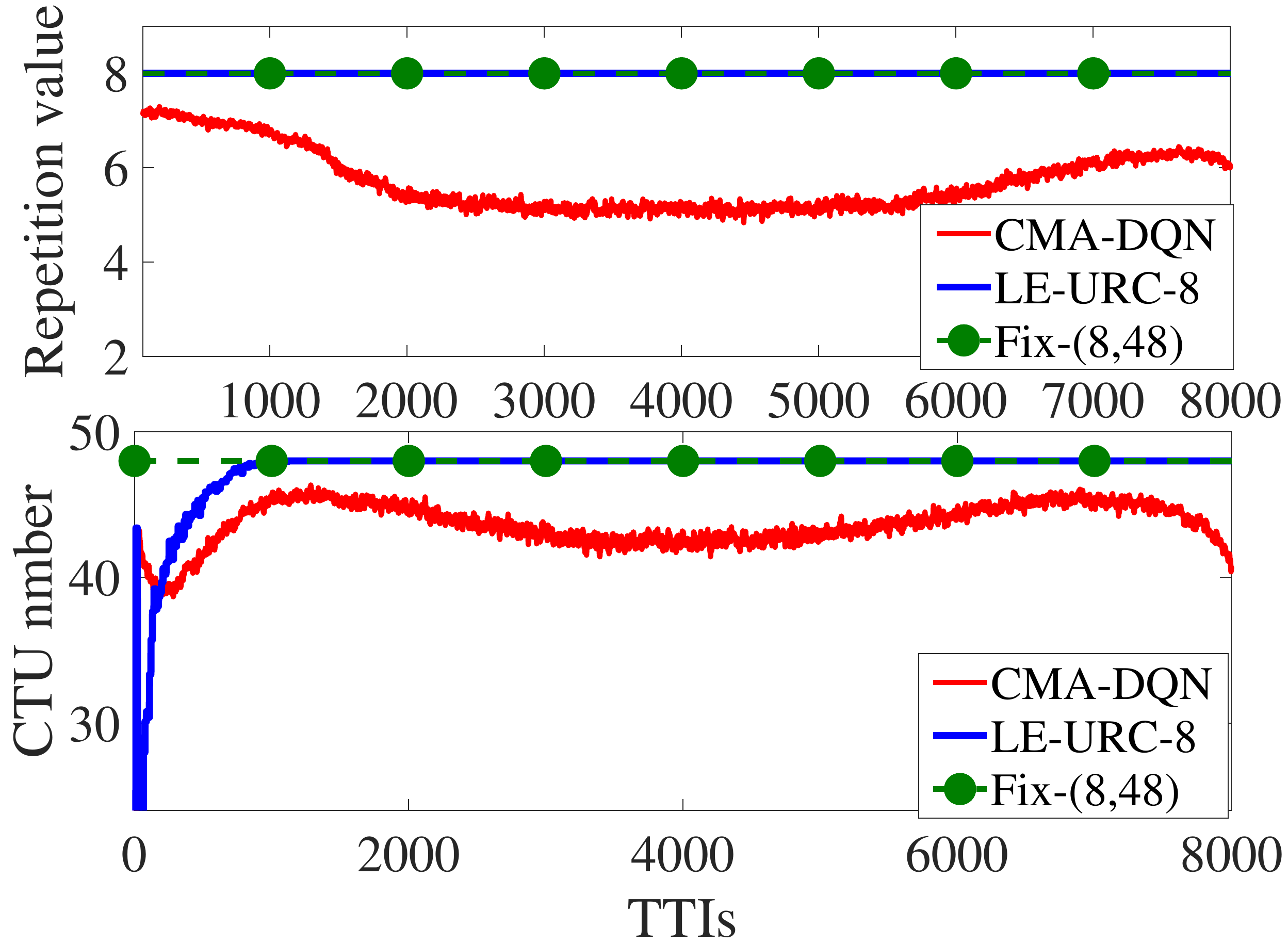}}
\subfigure[Proactive scheme]
{\includegraphics[width=2.6in,height=2.3in]{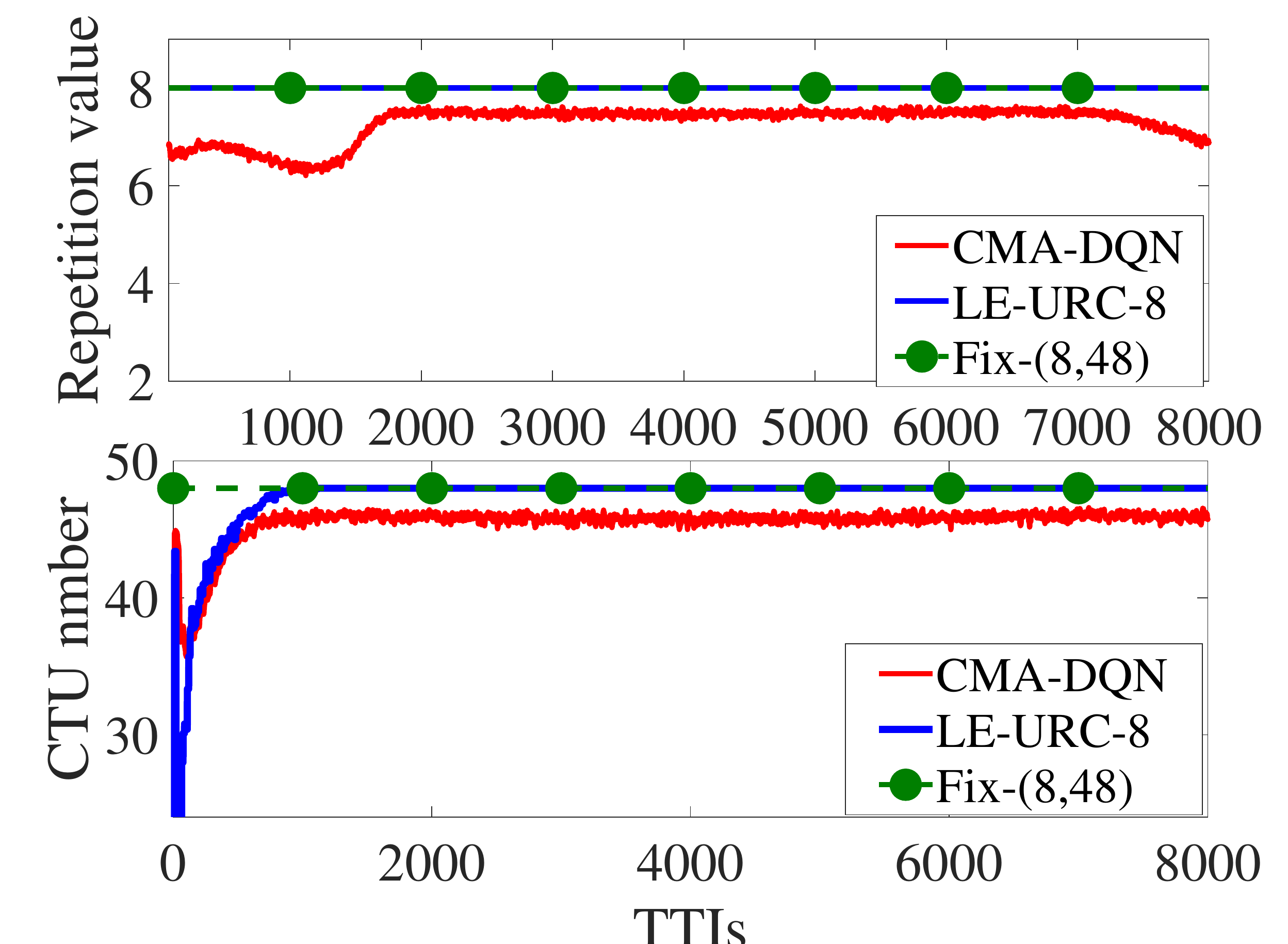}}
\caption{Actions for each GF scheme.}
\label{fig:15}
\end{figure}
Fig.~\ref{fig:15}  plots the average number $K$ of repetition values  and the average number $C$ of CTUs for each scheme that are selected
by CMA-DQN and LE-URC, respectively.
In Fig.~\ref{fig:15} (a), we can see that  the repetition value of K-repetition scheme decreases first and then increases back to a higher value.
This is because the agent in K-repetition scheme learns to sacrifice the current successful transmission to alleviate the traffic congestion in heavy traffic region to obtain a long-term reward, while LE-URC approach just adopts the maximum repetition value to optimize for one time instance.
In Fig.~\ref{fig:15} (b), we can see that the Proactive scheme adopts a higher and more stable repetition value due to its capability to deal with the traffic congestion.
In Fig.~\ref{fig:15} (a)(b), it can be seen that the number of CTUs has a similar trend as the repetition value, which may be caused by the sharing of actions as observations among agents.
% In Fig.~\ref{fig:15} (a), we can see that the Proactive scheme adopts a higher and more stable repetition value due to its capability to deal with the traffic congestion. However, the repetition value of K-repetition scheme decreases first and then increases back to a higher value. This is because the agent in K-repetition scheme learns to sacrifice the current successful transmission to alleviate the traffic congestion to obtain a long-term reward. In Fig.~\ref{fig:15} (b), It can be seen that the number of CTUs has a similar trend as the repetition value in Fig.~\ref{fig:15} (a), which may be caused by the sharing of actions as observations among agents.

% when the traffic is heavy, in our learning framework, the agent learns to dynamically configure lower repetition values and CTU numbers to alleviate the traffic congestion to obtain a long-term reward, while the LE-URC only optimizes at one time instance, which leads to a lower number of served UEs in long-term.

\section{Conclusion}
In this paper, we developed a general learning framework for dynamic resource configuration optimization in signature-based GF-NOMA systems, including the sparse code multiple access (SCMA), multiuser shared access (MUSA),  pattern division multiple access (PDMA), and etc,   for  mURLLC service under  the K-repetition GF scheme and the Proactive GF scheme.
This general learning framework was designed  to optimize the number of successfully served UEs under the latency constraint via adaptively configuring the uplink resources, including the repetition values and the contention-transmission unit (CTU) numbers.
We first performed a real-time repetition value configuration for the two schemes, where a  double  Deep  Q-Network  (DDQN)  was developed.  We then designed a Cooperative Multi-Agent learning technique based on the  DQN (CMA-DQN)  to optimize the configuration of both the repetition values and the CTU numbers for these two schemes, by dividing high-dimensional configurations into multiple parallel sub-tasks.
Our results have shown that:  1)  the number of successfully served UEs under the same latency constraint in our proposed learning framework is up to ten times for the K-repetition scheme, and two times for the Proactive scheme, more than that  with fixed repetition values and CTU numbers;  
2) with learning optimization, the  Proactive scheme always outperforms the K-repetition scheme in terms of the number of successfully served UEs, especially under the long latency constraint; 
3)  the proposed CMA-DQN is superior to the conventional load estimation-based approach  (LE-URC) that demonstrating its capability in dynamically configuring in long term;
% 2)  the superior performance of CMA-DQN over the conventional load estimation-based approach  (LE-URC) demonstrates its capability in dynamically configuring in long term;
4) the proposed learning framework is general and can be used to optimize the resource configuration problems in all the signature-based GF-NOMA schemes; and 5)  determining the retransmission or not can be optimized in the future by considering not only the long latency constraint but also the future traffic congestion, due to the fact that the long latency constraint will lead to high future traffic congestion.

%From the results using the simulated traffics in our work, we demonstrated that the proposed DRL-based resource configuration approach significantly outperform the networks with fixed transmission parameters
%in terms of throughput. 
%With realistic traffic, a direct implementation of DRL may bring computational
%complexity and processing delay at the NB-IoT BSs, so how
%to reduce the complexity of DRL algorithms can be considered
%in future work. 

\ifCLASSOPTIONcaptionsoff
  \newpage
\fi	

\bibliographystyle{IEEEtran}
\bibliography{IEEEabrv,grant}

\end{document}